\documentclass[a4paper]{aa}

\usepackage{graphicx}
\usepackage{txfonts}
\usepackage{lipsum}
\usepackage{subcaption}
\usepackage{placeins}
\usepackage{lscape}
\usepackage{wrapfig}

\usepackage[table]{xcolor}
\usepackage{colortbl}
\usepackage{url}

\definecolor{color1}{rgb}{0.965,1,1}
\newcolumntype{x}{>{\columncolor{color1}}c}
\definecolor{color2}{rgb}{1,0.965,1}
\newcolumntype{y}{>{\columncolor{color2}}c}
\definecolor{color3}{rgb}{1,1,0.965}
\newcolumntype{z}{>{\columncolor{color3}}c}

\definecolor{lightblue}{rgb}{0.96,0.98,1}
\newcommand{\bluerow}{\rowcolor{lightblue}}
\newcolumntype{b}{>{\columncolor{lightblue}}c}
\definecolor{lightgreen}{rgb}{0.98,1,0.98}
\newcommand{\greenrow}{\rowcolor{lightgreen}}
\newcolumntype{v}{>{\columncolor{lightgreen}}c}
\definecolor{lightgray}{gray}{0.97}
\newcommand{\grayrow}{\rowcolor{lightgray}}

\newcolumntype{g}{>{\columncolor{lightgray}}c}

\usepackage{caption}
\usepackage{float}
\usepackage{dblfloatfix}
\usepackage{rotating}

\usepackage{amsmath}
\usepackage{hyperref}
\usepackage[utf8]{inputenc}
\usepackage{multirow}

\begin{document} 

\title{Estimations for strong and weak hot Jupiter magnetism}
\subtitle{A Python package with interpolators and scaling laws}

    \authorrunning{Elias-López et al.}
    \titlerunning{Revised HJ magnetic field predictions}

   \author{Albert Elias-López \inst{1},
          Aline Vidotto \inst{1},
          Daniele Viganò \inst{2},
          J. R. Fuentes \inst{3},
          Simon Müller\inst{4},
          Emily Sandford \inst{1}}

    \institute {Leiden Observatory, Leiden University, PO Box 9513, 2300 RA Leiden, The Netherlands
    \and
    Institut de Ci\`encies de l'Espai (ICE-CSIC), Campus UAB, Carrer de Can Magrans s/n, 08193 Cerdanyola del Vallès, Barcelona, Catalonia, Spain
    \and
    TAPIR, California Institute of Technology, Pasadena, CA 91125, USA
    \and
    Department of Astrophysics, University of Zürich, Winterthurerstrasse 190, 8057 Zürich, Switzerland 
\\
\email{eliaslopez@strw.leidenuniv.nl}
             }

   \date{Received ---------; accepted --------}

    \abstract
   {Hot Jupiters (HJs) are commonly thought to host the strongest dynamo-generated magnetic fields among exoplanets, exceeding Jupiter's by up to an order of magnitude, which can potentially induce observable star-planet interaction (SPI) and planetary radio emission. However, while several indirect, model-dependent estimates of HJ magnetic fields have been reported from SPIs and recent transmission spectroscopy diagnostics, no unambiguous direct (radio) detection has yet been achieved.}
   {We aim to provide a range of magnetic field estimates for HJs, considering at the same time the mechanisms responsible for inflation and the convective flux that enters into the magnetic field scaling law.}
   {Given that their dynamos operate in the fast-rotating regime (Rossby number $\lesssim$ 0.1), we combine one-dimensional evolutionary interior models with magnetic field scaling laws under two limiting heating scenarios within the convective zone: deep and shallow energy depositions. Using interpolators built from the simulated data output, we provide magnetic field predictions for the HJ population.}
   {The deep-heating scenario maximizes convective heat flux in the dynamo and predicts strong magnetic fields of up to $\sim$100 G, whereas shallow heating, which is arguably more realistic for most proposed physical mechanisms, weakens convection and yields much weaker fields of ${\cal O}(1)$ G. The strong field scenario is compatible with SPI indirect detections, and the weak field scenario is in agreement with recent atmospheric constraints. We derive easy-to-use scaling law relations based on observables, i.e., planetary mass, radius, and/or incident stellar flux, that roughly reproduce the magnetic field estimates of our interior models: $B_{\rm strong} = 19 \, M_{\rm P}^{0.6} R_{\rm P}^{1.7}$ G; $B_{\rm weak} = 0.8 \, M_{\rm P}^{1.5} R_{\rm P}^{-1.2}$ G.}
   {With this range of magnetic field estimates for the current observed HJ population, our predictions provide a framework for interpreting future atmospheric wind measurements, SPI signatures, and radio observations, which may ultimately discriminate between competing interior-heating scenarios.}

   \keywords{Planets and satellites: magnetic fields - Planets and satellites: gaseous planets - Planets and satellites: interiors}

\maketitle
\nolinenumbers

\section{Introduction}\label{Sec: Introduction}

The magnetic field strength of astrophysical objects may depend on their rotation rate, depending on whether their internal dynamos operate in the slow- or fast-rotating regime. Observations of low-mass stars reveal a transition near a critical Rossby number of $\mathrm{Ro_{sat}}\sim0.13$, where $\mathrm{Ro}=P_{\rm rot}/\tau_{\rm turn}$, $P_{\rm rot}$ is the rotation period and $\tau_{\rm turn}$ the convective turnover time. This transition separates a rotation-dependent regime ($\mathrm{Ro} >\mathrm{Ro_{\rm sat}}$) from a saturated regime ($\mathrm{Ro} < \mathrm{Ro_{\rm sat}}$) in which magnetic activity no longer depends on rotation \citep[e.g.,][]{vilhu_nature_1984, noyes_rotation_1984, pizzolato_stellar_2003, wright_stellar-activity-rotation_2011, reiners_generalized_2014}. This low-$\mathrm{Ro}$ behavior is consistent with dynamo simulations and theory, which predict that magnetic field strength is governed primarily by the convective heat flux rather than the rotation rate \citep{christensen_energy_2009, davidson_scaling_2013}. These scaling relations successfully reproduce the magnetic field strengths of Earth, Jupiter, and low-mass stars, all of which host deep, rapidly rotating convective dynamos.

Hot Jupiters (HJs) are gas-giant exoplanets with orbital periods of a few days and separations $\lesssim0.1$ AU \citep{fortney_hot_2021} and are thus characterized by strong stellar irradiation and tidal interactions. One of their defining puzzles is the radius anomaly \citep{thorngren_hot_2024}: many HJs reach radii up to twice that of Jupiter, exceeding the predictions of standard evolutionary models. Although radius inflation correlates with stellar irradiation once $T_{\rm eq}\gtrsim1000$ K \citep{weiss_mass_2013}, irradiation alone cannot reproduce radii above $\sim1.3 \, \mathrm{R_J}$ at Gyr ages \citep{guillot_giant_1996, arras_thermal_2006, burrows_possible_2007, fortney_planetary_2007}. The excess inflation is therefore generally attributed to an additional heat source linked to the incident stellar irradiation \citep[see][for reviews]{fortney_hot_2021, thorngren_hot_2024}.

HJs are generally expected to be tidally locked, rotating with periods of a few days rather than a typical Jovian value of $\sim10$ h for cold giants \citep{batygin_terminal_2018}. Nevertheless, recent Rossby number estimates indicate that their deep convective interiors remain in the rapidly rotating regime (${\rm Ro}<{\rm Ro}_{\rm sat}$), where dynamo scaling laws are expected to apply \citep{elias-lopez_rossby_2025}. These scaling laws predict magnetic fields more than one order of magnitude stronger than Jupiter's if the extra heat driving inflation enhances convection within the dynamo region \citep{yadav_estimating_2017}. Conversely, if the heat is deposited in the outer convective layers, as expected in Ohmic dissipation models \citep{batygin_evolution_2011, batygin_magnetically_2013, knierim_shallowness_2022, vigano_inflated_2025}, the predicted magnetic fields may be even weaker than Jupiter's \citep{elias-lopez_rossby_2025}.

Because dynamo scaling laws predict stronger magnetic fields with increasing planetary mass and internal luminosity, inflated HJs are promising targets for searches for Jovian-like coherent radio emission \citep[e.g.,][]{farrell_possibility_1999}. Such emission, observed from all magnetized Solar-System planets \citep{zarka_auroral_1998}, is detectable from the ground only if the electron cyclotron frequency, $\nu_c\simeq2.8\,B$ [G] MHz, exceeds the $\sim10$ MHz ionospheric cutoff and the radio flux is sufficiently high. The latter remains highly uncertain because predictions are largely calibrated to Solar-System planets \citep[e.g.,][]{griesmeier_predicting_2007}. Despite numerous searches and a growing population of radio-loud stars \citep{callingham_population_2021, yiu_radio_2024, pritchard_multi-epoch_2024, driessen_sydney_2024}, no unambiguous planetary radio emission has yet been detected \citep[e.g.,][]{gasperin_radio_2020, turner_search_2021, cordun_deep_2025}. Independent constraints come from star--planet interactions (SPIs), including orbital modulation of Ca {\sc II} K and high-energy emission \citep{shkolnik_evidence_2003, pillitteri_fuv_2015, cauley_evidence_2018}. From such signatures, \citet{cauley_magnetic_2019} inferred surface fields of tens to $\sim100$ G for four HJs, although these estimates are strongly model-dependent. In contrast, atmospheric wind measurements for seven HJs suggest magnetic fields of at most a few gauss to avoid excessive magnetic drag \citep{seidel_magnetic_2026}.

Motivated by these contrasting observational constraints and the relative scarcity of available theoretical predictions, we investigate how different internal heating scenarios affect the magnetic field predictions for HJs. Using one-dimensional planetary evolution models combined with a dynamo scaling law for rapidly rotating objects, we estimate magnetic fields for two limiting heating prescriptions that correspond to energy deposition either inside or outside the dynamo region. From these models, we construct an interpolator and derive analytical scaling laws that estimate magnetic fields from the observed masses, radii, and incident stellar fluxes of HJs. These publicly available tools (see Code availability) provide practical resources for interpreting current observations and guiding future observational studies. The paper is organized as follows. In Sect.~\ref{Sec: Method}, we describe the planetary evolution models, the two heating prescriptions, and the methodology used to estimate the magnetic field strengths. In Sect.~\ref{Sec: Results}, we present magnetic field estimates for the observed HJ population and derive analytical scaling laws suitable for quick estimates. In Sect.~\ref{Sec: Discussion}, we discuss how the model assumptions might influence our results, and compare our predictions with current indirect estimates of HJ magnetic fields. Finally, we summarize our conclusions in Sect.~\ref{Sec: Conclusions}.

\section{Method}\label{Sec: Method}

\subsection{Evolutionary models}\label{Sec: Evolutionary model}

To numerically model the interiors of inflated HJs, we follow \cite{elias-lopez_rossby_2025}: we use the public stellar evolution code MESA\footnote{\url{https://mesastar.org/}} (version 24.08.1) to evolve highly irradiated and internally heated Jupiter-like planets \citep{paxton_modules_2011, paxton_modules_2013, paxton_modules_2015, paxton_modules_2018, paxton_modules_2019, jermyn_modules_2023}. We solve the one-dimensional time-dependent stellar structure equations, that is, mass conservation, hydrostatic equilibrium, energy conservation, and energy transport, which are, respectively:
\begin{gather}
    \frac{dr}{dm} = \frac{1}{4 \pi r^2~\rho} ~,
    \label{Eq: MESA mass conservation}\\
    \frac{dP}{dm} = - \frac{G m}{4\pi r^4} ~,
    \label{Eq: MESA hydrostatic equilibrium}\\
    \frac{dL}{dm} = - T \frac{ds}{dt} + q_{\rm irr}(m) + q_{\rm heat}(m) ~,
    \label{Eq: MESA luminosity}\\
    \frac{dT}{dm} = - \frac{G m T}{4\pi r^4 P} \nabla ~,
    \label{Eq: MESA energy transport}
\end{gather}
where $m$ is the mass enclosed within a radius $r$, $\rho$ is the density, $P$ is the pressure, $G$ the gravitational constant, $s$ the specific entropy, $T$ the temperature, $L$ the internal luminosity, and $\nabla \equiv d \ln T / d \ln P$ is the logarithmic temperature gradient. The energy equation (Eq.~\ref{Eq: MESA luminosity}) includes the specific heating and cooling rates on the right-hand side (in units of erg g$^{-1}$ s$^{-1}$). They are, in order, the gravitational contraction, stellar irradiation $q_{\rm irr}$, and a general internal heat source $q_{\rm heat}$. The latter has been previously used to model the effect of continuous heat deposition, and thus inflation, in HJs \citep[e.g.,][]{komacek_structure_2017, thorngren_bayesian_2018, elias-lopez_rossby_2025, vigano_inflated_2025}. The set of equations is closed using the MESA equation of state (EoS), which, for the density and temperature regimes of gas-giant and brown-dwarf interiors, primarily interpolates the \cite{saumon_equation_1995} EoS for H-He mixtures. For all our models, we assume a 10 $M_{\oplus}$ inert homogeneous rocky core with density $\rho_c=10$~g~cm$^{-3}$, thus fixing the core size $R_{\rm core}$. If not stated otherwise, the envelope has a fixed solar composition, i.e., $(X, Y, Z)=(0.74, 0.24, 0.02)$. We also explore the influence of more refined EoS and different envelope metallicities (see Sect.~\ref{Sec: Discussion} and App.~\ref{App: Interior model caveats}).

For simplicity, stellar irradiation is modeled as atmospheric heat injection deposited uniformly in mass throughout the outermost mass column, that is, when $m(r) \geq M_{\rm P} - 4 \pi R_{\rm P}^2 \Sigma_{\rm irr}$. The absorption column depth, $\Sigma_{\rm irr}$, is fixed at 200 g cm$^{-2}$, similar to previous works \citep{komacek_structure_2017, vigano_inflated_2025}. It approximately corresponds to a gray opacity $\kappa=5\times 10^{-3}$~cm$^2$~g$^{-1}$, with a typical maximum depth of a fraction of a bar, which corresponds to a certain radius, denoted as $R_{\rm irr}$, at the base of the irradiated layer. The specific energy absorption rate is then $q_{\rm irr}(m) = F_{\rm irr} / (4 \Sigma_{\rm irr})$, where the received stellar irradiation is $F_{\rm irr} =L_{\star}/(4 \pi a^2) = \sigma_{\rm SB} T_{\rm eff}^4 \,R_{\star}^2/a^2$. For simplicity, we assume $F_{\rm irr}$ is constant in time, neglecting its gradual increase during main-sequence stellar evolution (see e.g. \citealt{komacek_reinflation_2020, vigano_inflated_2025} for such effects).

The physical mechanism responsible for HJ inflation remains uncertain, with proposed explanations involving either reduced internal cooling or additional energy deposition within the planet \citep[see the review by][]{fortney_hot_2021}. The heat deposition profile, $q_{\rm heat}(m)$, strongly affects the extent of the convective regions \citep[e.g.,][]{ginzburg_hot-jupiter_2015, ginzburg_extended_2016, komacek_structure_2017}, and is therefore particularly relevant for magnetic-field estimates based on the scaling law of \citet{christensen_energy_2009}. This is because the heat required to explain the inflated radii of HJs can exceed the planet's intrinsic cooling luminosity \citep{thorngren_bayesian_2018, yadav_estimating_2017}, making the predicted magnetic field sensitive to the heating prescription and its resulting convective structure \citep{elias-lopez_rossby_2025}. Rather than adopting a specific inflation mechanism, we consider two idealized limiting heating prescriptions that isolate the effect of the deposition depth on the planetary structure and magnetic field (see Sect.~\ref{Sec: Conclusions} for a further discussion).

\begin{figure}[t]
\centering
\includegraphics[width=.9\hsize]{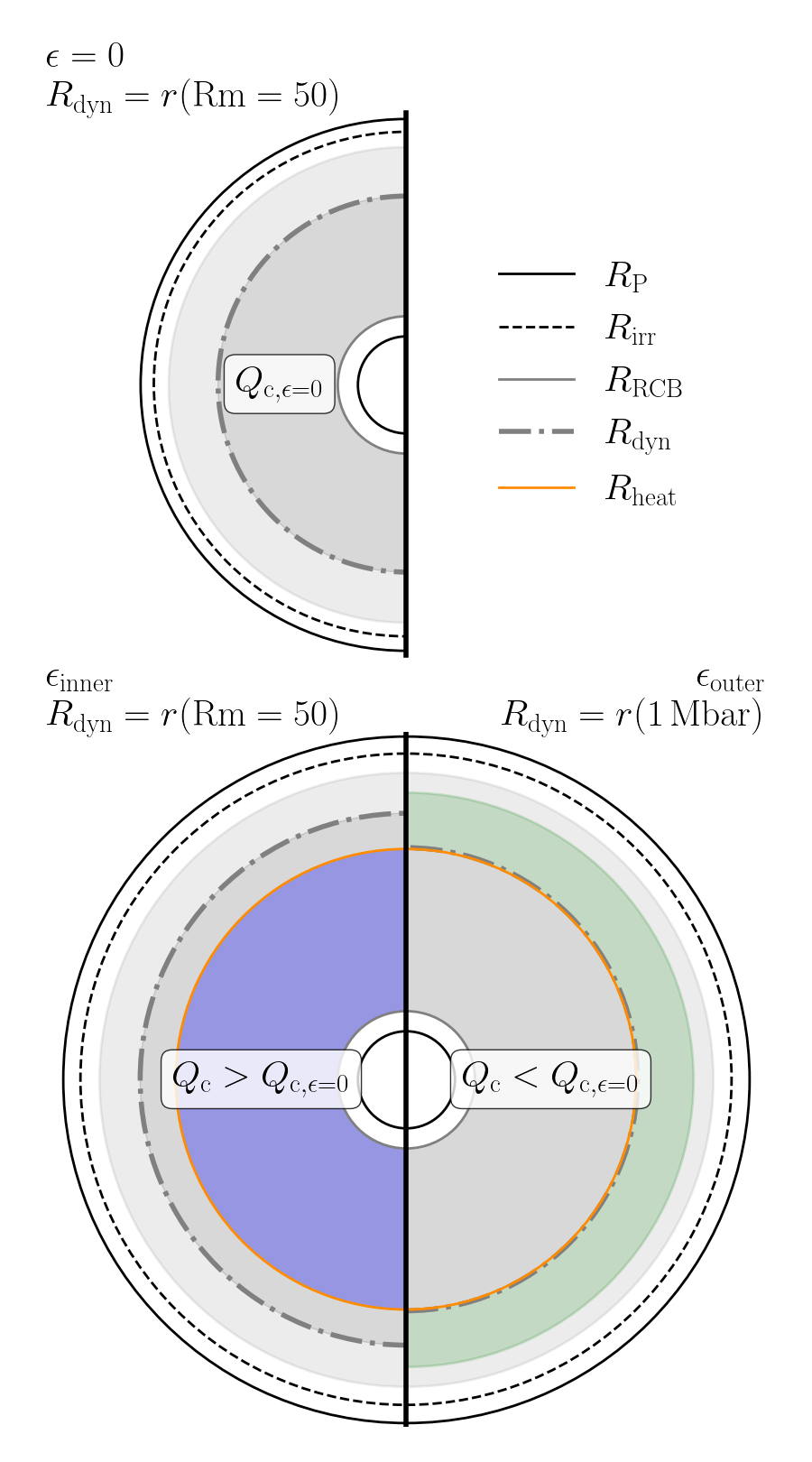}
\caption{Schematic illustration of the planetary models; the radial scales and relative thicknesses of the different layers are not to scale and are chosen for illustrative purposes. Top: unheated case ($\epsilon=0$), with an intrinsic cooling luminosity denoted by $Q_{c,\epsilon=0}$. Bottom: internally heated models, $\epsilon_{\rm inner}$ (left) and $\epsilon_{\rm outer}$ (right). The blue (green) region indicates the location of heat deposition for $\epsilon_{\rm inner}$ ($\epsilon_{\rm outer}$), resulting in $Q_{\rm c}>Q_{c,\epsilon=0}$ ($Q_{\rm c}<Q_{c,\epsilon=0}$). The colored areas indicate the convective regions. For $\epsilon_{\rm outer}$, additional stably stratified layers may develop in the deep interior, depending on the planetary mass and the amount of injected heat. The radial scales shown in the figure are the planet radius $R_{\rm P}$, the irradiation depth $R_{\rm irr}$, the radiative–convective boundary $R_{\rm RCB}$, the dynamo radius  $R_{\rm dyn}$ (see Sect.~\ref{Sec: Dynamo radius definition}), and the heating boundary $R_{\rm heat}$.}

\label{Fig: HJ diagrams}
\end{figure}

\begin{figure}[t]
\centering
\includegraphics[width=\hsize]{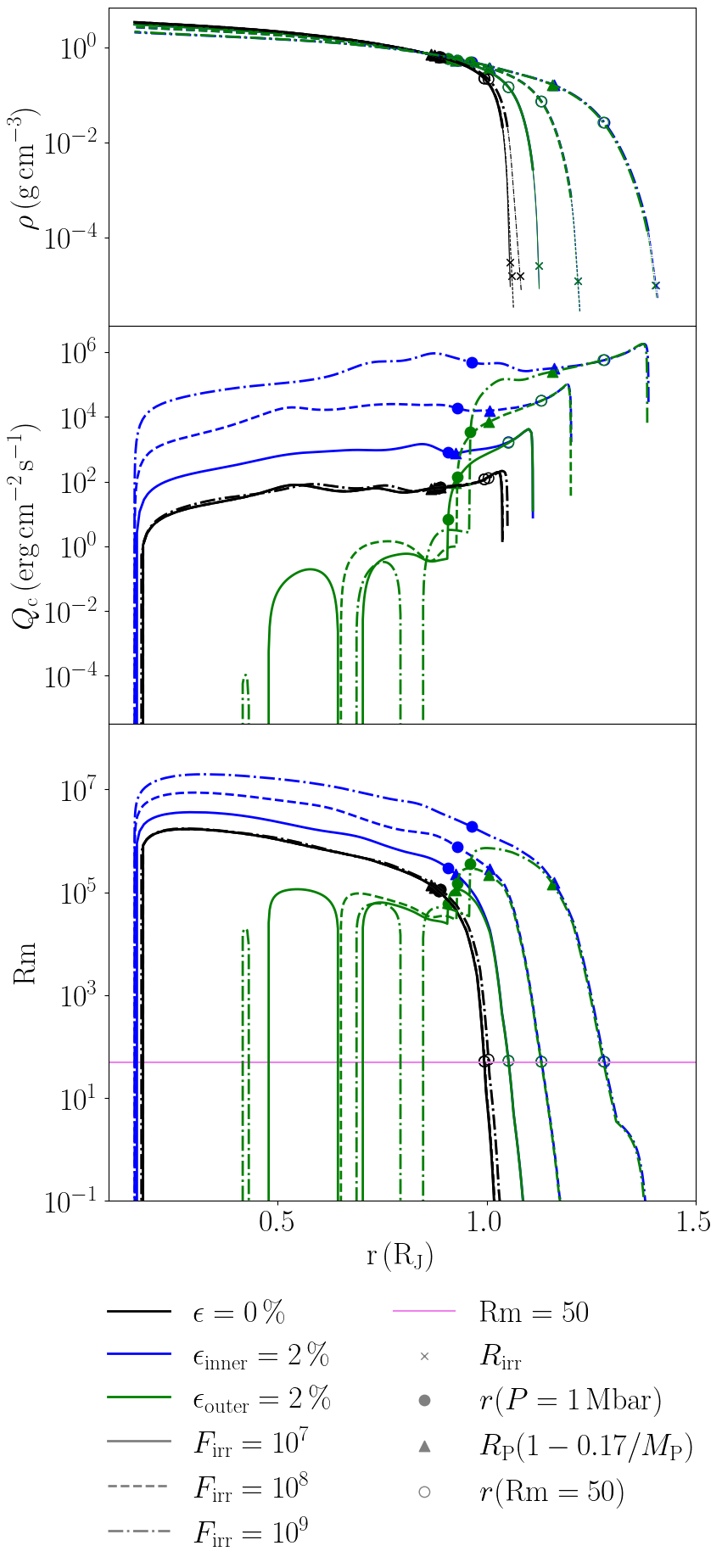}
\caption{Interior radial profiles of the density (top), convective heat flux (middle; Eq.~\ref{Eq: Convective heat flux}), and magnetic Reynolds number (bottom; Eq.~\ref{Eq: Rm}) for irradiated-only ($\epsilon=0$; black) and internally heated ($\epsilon_{\rm inner}$ and $\epsilon_{\rm outer}$; blue and green) models. Different line styles denote different incident stellar fluxes, $F_{\rm irr}$ (in $\mathrm{erg\,cm^{-2}\,s^{-1}}$). Thick and thin lines show convective and radiative regions, respectively. Crosses mark the base of the stellar irradiation layer, while the remaining symbols indicate the different $R_{\rm dyn}$ definitions (see Sect.~\ref{Sec: Dynamo radius definition}).}
\label{Fig: HJ profiles}
\end{figure}

The total deposited power, $Q_{\rm dep}=\int_{M_{\rm shell}}q_{\rm heat}(m)\,{\rm d}m$, is quantified by the heating efficiency $\epsilon=Q_{\rm dep}/(\pi R^2F_{\rm irr})$ defined as the ratio between the deposited heat and the incident stellar flux intercepted by the planet. For simplicity, we assume a uniform specific heating rate, $q_{\rm heat}=Q_{\rm dep}/M_{\rm shell}$, within a shell of mass $M_{\rm shell}$ bounded by $r_{\rm bot}$ and $r_{\rm top}$. We consider two limiting cases, either above or below a specific radius, $R_{\rm heat}=r(P=1\,{\rm Mbar})$, which is of the order of the dynamo radius (i.e., $R_{\rm dyn}$, see Sect.~\ref{Sec: Dynamo radius definition}):
\begin{enumerate}
    \item Inner deposition ($\epsilon_{\rm inner}$): heat is deposited from $r_{\rm bot}=R_{\rm core}$ and $r_{\rm top}=R_{\rm heat}$.
    \item Outer deposition ($\epsilon_{\rm outer}$): heat is injected from $r_{\rm bot}=R_{\rm heat}$ and $r_{\rm top}\leq R_{\rm RCB}$, where $R_{\rm RCB}$ is the radiative--convective boundary. 
\end{enumerate}
Note that, in both cases, no extra heat is deposited in the radiative region because it would not efficiently inflate the planet. Representative radial profiles for the $\epsilon_{\rm inner}$, $\epsilon_{\rm outer}$, and non-inflated ($\epsilon=0$) models are shown in Fig.~\ref{Fig: HJ profiles}, illustrating their different density and convective structures. Figure~\ref{Fig: HJ diagrams} schematically illustrates the resulting structures of the unheated and internally heated planetary models.

The inner-heating prescription maximizes the heat transported through the dynamo and therefore the predicted magnetic field strengths \citep{elias-lopez_rossby_2025}. It is conceptually equivalent to the work of \cite{yadav_estimating_2017}, which assumes the inflation heat corresponds to the dynamo thermal luminosity. Conversely, the outer-heating prescription minimizes the predicted magnetic field by excluding the deposited heat from the dynamo heat flux and reducing the total convective power. This is motivated by most proposed inflation mechanisms depositing energy primarily in the low-density envelope outside the metallic dynamo region \citep[e.g.,][]{batygin_inflating_2010, batygin_evolution_2011, wu_ohmic_2012, ginzburg_hot-jupiter_2015, ginzburg_extended_2016}. For example, Ohmic dissipation predicts a heating profile approximately proportional to $\rho^{-2}$, concentrating energy in the outer layers with negligible heating below the 1 Mbar level \citep{komacek_structure_2017, knierim_shallowness_2022, vigano_inflated_2025}. As shown by \citet{elias-lopez_rossby_2025}, a sufficiently strong $\epsilon_{\rm outer}$ can substantially modify the internal convective structure by producing stably stratified layers, where the adiabatic gradient surpasses the radiative gradient ($\nabla_{\rm ad}>\nabla_{\rm rad}$) and thus convection is suppressed ($v_{\rm conv}=0$, $Q_{\rm c}=0$, and $\mathrm{Rm}=0$; see Sects.~\ref{Sec: Convective velocities} and \ref{Sec: Magnetic field scaling laws} for definitions). 

Heating efficiencies of only a few percent at most are sufficient to reproduce the observed HJ radius distribution \citep[e.g.,][]{batygin_inflating_2010, batygin_evolution_2011, wu_ohmic_2012, komacek_structure_2017}. Statistical studies further indicate that $\epsilon$ peaks at $\sim2.4\%$ for $T_{\rm eq}\approx1500$--1900 K before declining at higher equilibrium temperatures \citep{thorngren_bayesian_2018, sarkis_evidence_2021}.

\subsection{Magnetic field scaling law}\label{Sec: Magnetic field scaling laws}

Due to the strong tidal interactions with their host star, HJs are usually assumed to be tidally locked. Despite this, their convective interiors were shown to remain in the fast-rotator regime (i.e., $\mathrm{Ro}\,< 0.1$) in the regions where metallic hydrogen is believed to drive dynamo action \citep{elias-lopez_rossby_2025}. Therefore, we can assume the magnetic field scaling law introduced by \cite{christensen_scaling_2006, christensen_energy_2009} holds, which states that for convective fast-rotating objects, the convective heat flux determines the dynamo-generated magnetic field strength: $B^2/2 \mu_0 \propto Q_{\rm c}^{2/3}$. This expression can be integrated in the radial direction (between $R_{\rm core}$ and $R_{\rm dyn}$) to calculate the mean magnetic field strength at the dynamo surface:
\begin{equation}
\begin{aligned}
    \frac{B_{\rm rms, dyn}^2}{2 \mu_0} = c \, f_{\rm ohm} \, \langle~\rho~\rangle^{1/3} \, \left( F \, Q_0 \right)^{2/3}, \quad \text{where} \\ F^{2/3} = \frac{1}{V} \int_{R_{\rm core}}^{R_{\rm dyn}} \left( \frac{Q_{\rm c}(r)}{Q_0} \frac{L(r)}{H_T(r)}~\right)^{2/3} \left( \frac{\rho(r)}{\langle~\rho~\rangle}~\right)^{1/3} 4 \pi r^2 \, dr
\end{aligned}
\label{Eq: B scaling law}
\end{equation}
where $\mu_0$ is the magnetic permeability, $\langle\rho\rangle$ is the volume-averaged density, $c=0.63$ is an order-unity calibration constant, and $f_{\rm ohm}\leq1$ is the Ohmic dissipation ratio, which we assume to be unity in agreement with the geomagnetic, Jovian, classical T Tauri stars, old M dwarfs, and rapidly rotating solar-type stars \citep[i.e., $P_{\rm rot}<4$ days,][]{christensen_energy_2009}. The largest convective length scale is taken as $L=\min(D, H_P)$, where $D=R_{\rm dyn}-R_{\rm core}$ is the thickness of the convective dynamo region and $H_P$ is the pressure scale height. The temperature scale height is defined as $H_T=P/(\rho g\nabla_{\rm ad})$, where $g$ is the gravitational acceleration. Finally, $Q_{\rm c}$ is the convective heat flux and $Q_0$ is a reference heat flux used to normalize the integral. This implementation follows previous studies of planetary dynamos \citep[e.g.,][]{elias-lopez_rossby_2025,hori_linkage_2021,kilmetis_magnetic_2024}. Since the integral depends on the convective heat flux throughout the dynamo region, it is higher for the $\epsilon_{\rm inner}$ heating prescription than the $\epsilon_{\rm outer}$ one. 

\cite{reiners_surprisingly_2009} proposed a simplified version of Eq.~\ref{Eq: B scaling law}, expressing the magnetic field strength solely as a power law of the mass, radius, and luminosity of the astrophysical object. Hereafter, we refer to this scaling law as R09. In App.~\ref{App: R09 scaling law}, we present its full formulation and compare it with the results derived in this work. Additionally, we estimate the dipolar component at the planetary surface at the magnetic equator, $B_{\rm dip, surf}$, defined as: 
\begin{equation}
    B_{\rm dip, surf} = K_{\rm dip} \left( \frac{R_{\rm dyn}}{R_{\rm P}}\right)^3 B_{\rm rms, dyn}~,
    \label{Eq: Bdip}
\end{equation}
where we take $K_{\rm dip}=1/\sqrt{2}$ to account for the relation between the rms dynamo field and its dipolar component (see App.~\ref{App: Rotation correction and scaling law calibration} for the calibration and associated caveats).

As mentioned above, sufficiently strong $\epsilon_{\rm outer}$ can produce stably stratified layers that divide the convective region into multiple thinner shells, for which the assumptions of a single convection-driven dynamo are not strictly satisfied. Numerical dynamo studies show that shell thickness can affect the dynamo regime and magnetic field morphology \citep[e.g.,][]{heimpel_numerical_2005}, with thin-shell dynamos exhibiting similarities to stratified anelastic dynamos, where weak multipolar solutions can be favored \citep{gastine_dipolar_2012, duarte_anelastic_2013}. Nevertheless, the magnetic-field scaling with available convective power remains broadly consistent with Boussinesq results across a range of shell thicknesses and density stratifications, despite differences in field morphology \citep{yadav_consistent_2013, schrinner_topology_2014}. Moreover, the geometry of these possible stratified layers is dependent on the heating profiles \citep{ginzburg_extended_2016, komacek_structure_2017} and may also be sensitive to interior model assumptions, such as the EoS, metallicity, and mixing-length treatment. We therefore do not regard the detailed geometry of the stratified and convective layers predicted by the $\epsilon_{\rm outer}$ models as robust enough to motivate a more specific dynamo treatment, for which physically motivated heating profiles would be required.

Although Eq.~\ref{Eq: B scaling law} is not strictly validated for the $\epsilon_{\rm outer}$ models with thin or separated convective shells, we apply it as an order-of-magnitude estimate of the dynamo field strength. Similarly, our conversion from the rms dynamo field to the surface dipole assumes a dipole-dominated dynamo (Eq.~\ref{Eq: Bdip}). Recent simulations show that multipolar dynamos can have dipole components approximately an order of magnitude weaker than dipolar dynamos at comparable driving power \citep{wulff_new_2026}. Thus, for models with thin or separated convective shells ($\epsilon_{\rm outer}$), the inferred $B_{\rm dip,surf}$ should be regarded as an upper estimate of the dipolar field strength (see Sect.~\ref{Sec: Conclusions} for the multipolar case).

\subsection{Heat flux and convective velocity}\label{Sec: Convective velocities}

To calculate the convective heat flux used in Eq.~\ref{Eq: B scaling law}, we follow \citet{hori_linkage_2021} and adopt
\begin{equation}
    Q_{\rm c} = \frac{4 \, c_P \, T \,~\rho^2 \, v_{\rm conv}^3}{\alpha_{\rm MLT} P \, \delta}
    \label{Eq: Convective heat flux}
\end{equation}
where $c_P$ is the heat capacity, $\delta = - (\partial \, \mathrm{ln} \rho/ \partial \, \mathrm{ln} \, T)_P$, and $v_{\rm conv}$ is the convective velocity. The latter is calculated via the MESA default scheme; that is, it is derived from mixing-length theory (MLT) based on the model in \cite{kuhfuss_model_1986}, which reduces to \cite[chap. 14]{cox_principles_1968} in the limit of long time steps (see \citealt{paxton_modules_2011, jermyn_modules_2023} for details). As in \citet{hori_linkage_2021}, we adopt $\alpha_{\rm MLT}=2$.

A novel aspect of this work is the inclusion of a rotation factor that diminishes the convective velocity, $v_{\rm conv}$ \citep{fuentes_rotation_2023}. Standard mixing-length theory estimates the convective velocity by balancing buoyancy and inertia (the free-fall velocity), but this approximation breaks down in rapidly rotating objects such as Jupiter, Saturn, and HJs. In the quasi-geostrophic regime, convection is instead governed by a balance between Coriolis, inertial, and buoyancy forces (the CIA balance; \citealt{stevenson_turbulent_1979, fuentes_rotation_2023}). Following \citet{fuentes_rotation_2023} and \citet{hindman_dwindling_2023}, we therefore reduce the non-rotating convective velocity, $v_{\rm conv,NR}$, according to
\begin{equation}
    v_{\rm conv, R} = v_{\rm conv, NR} \left( \frac{v_{\rm conv, NR}}{2 \Omega H_P} \right)^{1/5}
    \label{Eq: Reduced convective velocity}
\end{equation}
where $\Omega=2\pi/P_{\rm orb}$ is the planetary rotation rate, assuming tidal synchronization. This scaling has recently been supported by three-dimensional hydrodynamical simulations employing both fixed and time-dependent heat fluxes \citep[see Fig.~3 of][]{zhang_3d_2025}. For Jupiter, this factor corresponds to a convective velocity reduction by a factor of $\sim$6, increasing the convective turnover time from $\sim$1 to $\sim$5 years. The convective heat fluxes shown in Fig.~\ref{Fig: HJ profiles} are computed using Eqs.~\ref{Eq: Convective heat flux} and \ref{Eq: Reduced convective velocity}. We apply the rotational correction as a post-processing step after computing the interior structure, so its feedback on the planetary evolution is not included. The impact of this correction on the predicted magnetic field strengths is discussed in App.~\ref{App: Rotation correction and scaling law calibration}.

\begin{figure*}[th!]\centering
\includegraphics[width=.42\textwidth]{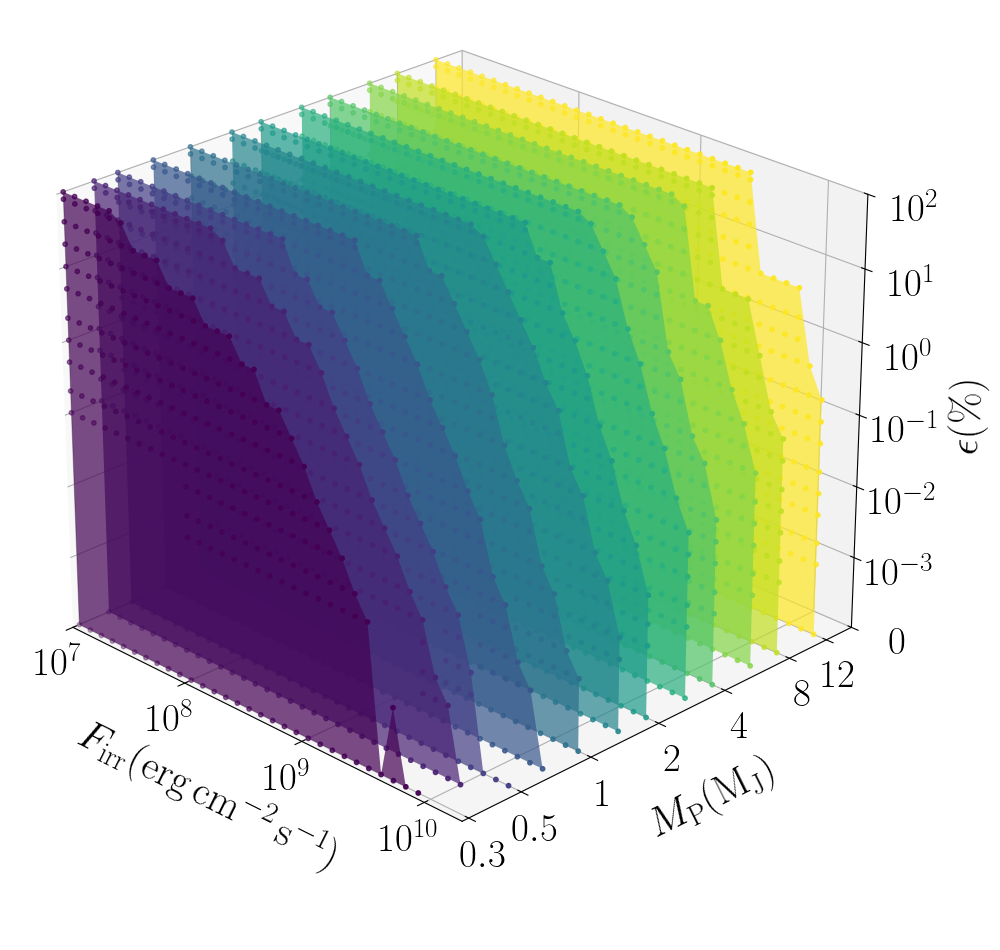}\includegraphics[width=.55\textwidth]{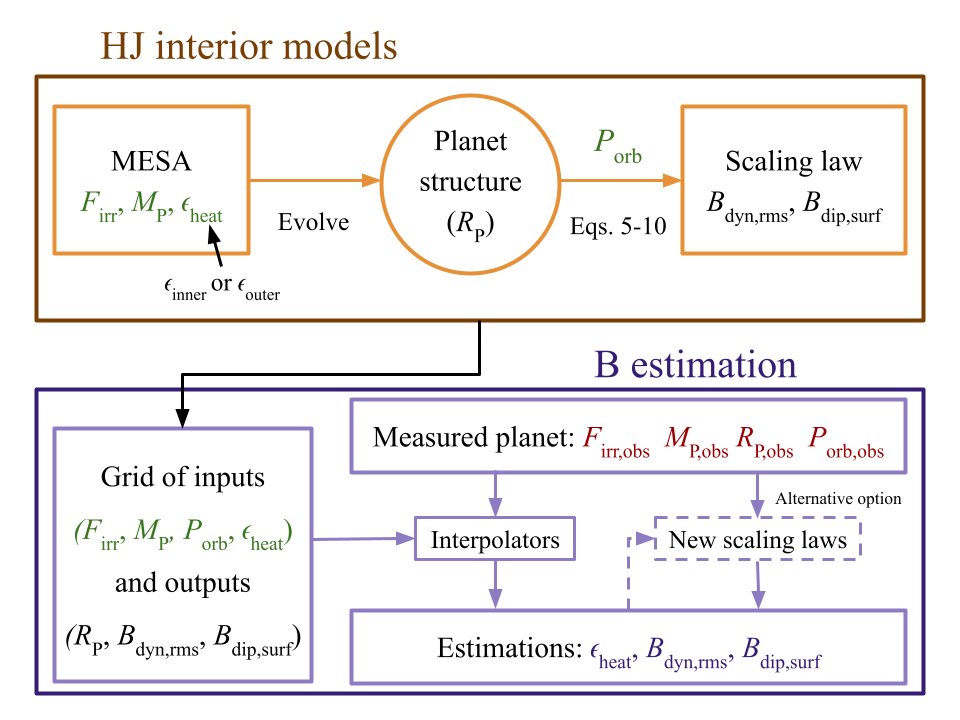}
\caption{Left: grid of input parameters used in our evolutionary models. Each colored slice corresponds to models with different planetary masses. Right: schematic overview of our modeling framework. For each combination of incident stellar flux, planetary mass, orbital period, and heating efficiency (quantities in green), we compute planetary interior models and estimate the corresponding dynamo and surface dipolar magnetic fields. These models are then used to construct interpolation grids that estimate the heating efficiency and magnetic field strengths of observed HJs from their measured irradiation, mass, radius, and orbital period (quantities in red). Approximate analytical scaling laws provide a faster, albeit less accurate, alternative.}
\label{Fig: Simulation grid and method}
\end{figure*}

\subsection{Dynamo radius definition}\label{Sec: Dynamo radius definition}

The dynamo scaling law presented in Sect.~\ref{Sec: Magnetic field scaling laws} requires defining the outer boundary of the dynamo region, $R_{\rm dyn}$, as this determines both the volume over which the convective power is integrated (Eq.~\ref{Eq: B scaling law}) and the attenuation of the magnetic field between the dynamo and the planetary surface (Eq.~\ref{Eq: Bdip}). Three commonly used choices can be considered: (i) an analytical expression calibrated to Jupiter \citep{reiners_magnetic_2010}, (ii) the radius corresponding to a pressure of 1 Mbar, where hydrogen becomes metallic \citep{french_ab_2012, bonitz_toward_2024}, and (iii) the radius at which the magnetic Reynolds number reaches the critical value required for dynamo action (or, equivalently, where the electrical conductivity becomes sufficiently high). We therefore first examine these definitions and then adopt appropriate dynamo radii for our two heating scenarios.

Numerical dynamo simulations of fully convective thick shells show that magnetic field generation requires the magnetic Reynolds number to exceed a critical value of order $\mathrm{Rm}_{\rm crit}\sim50$ \citep{christensen_scaling_2006}, where
\begin{equation}
    \mathrm{Rm}=\frac{v_{\rm conv}H_P}{\eta}
    =\mu_0\sigma v_{\rm conv}H_P,
    \label{Eq: Rm}
\end{equation}
with $H_P$ the pressure scale height (adopted as the characteristic length scale), $\eta$ the magnetic diffusivity, and $\sigma$ the electrical conductivity. Typical values in the metallic hydrogen region of gas giants are $\mathrm{Rm}\sim10^6$--$10^7$, well above the dynamo threshold\footnote{Assuming Jovian values inside the metallic hydrogen region, i.e., $H_P\sim D\sim5\times10^7$ m, $v_{\rm conv}\sim0.1$ m s$^{-1}$ and $\sigma\sim10^6$ S m$^{-1}$.}. Outside this highly conducting region, $\sigma$ decreases by several orders of magnitude and $\mathrm{Rm}$ eventually falls below the critical value.

To determine where this transition occurs, we compute $\sigma$ profiles by combining state-of-the-art conductivity tables for both the shallow and deep interior, following the methodology of \cite{vigano_inflated_2025}.\footnote{The tables and interpolation routine are available at \url{https://github.com/danielevigano/mesa_ohmic_hj}.} Briefly, we interpolate in the $\log T$-$\log\rho$ plane using the alkali thermal ionization tables of \cite{kumar_ionization_2021}, dominant at $\rho\lesssim10^{-2}$ g cm$^{-3}$, and the pressure-ionized H-He conductivity tables of \cite{holst_electronic_2011, bonitz_toward_2024}, dominant at higher densities. The bottom panel of Fig.~\ref{Fig: HJ profiles} shows the resulting $\mathrm{Rm}(r)$ profiles for models with different irradiation levels and internal heating rates.

The value $\mathrm{Rm}_{\rm crit}\simeq50$ should not be interpreted as a sharp or universal boundary: it was derived from global magnetic Reynolds numbers in numerical simulations under the Boussinesq approximation \citep{christensen_scaling_2006}, whereas here we evaluate a local $\mathrm{Rm}(r)$. It has been shown that the critical value can depend on the dynamo properties and may be larger than 50 \citep[e.g.,][]{schaeffer_energetically_2025}. Unlike non-irradiated gas giants, HJs retain a partially ionized outer envelope due to thermal ionization of alkali metals \citep{kumar_ionization_2021, dietrich_magnetic_2022}. This smooths out the conductivity profile and increases the region showing a large intermediate dynamo regime (i.e., between the threshold for metallic hydrogen at 1 Mbar and the threshold for $\mathrm{Rm}=50$), compared to a cold Jupiter, which instead shows an exponential drop in conductivity. As shown in the bottom panel of Fig.~\ref{Fig: HJ profiles}, this region increases with planetary inflation, since hotter planets have both higher thermal ionization and larger heat deposition. 

We adopt two limiting definitions of $R_{\rm dyn}$ for the different heating mechanisms:
\begin{enumerate}
    \item For the $\epsilon_{\rm inner}$ models we adopt $R_{\rm dyn}=r(\mathrm{Rm}=50)$. These points are marked by open circles in Fig.~\ref{Fig: HJ profiles}. Since the dynamo extends farther out in the planet, this choice maximizes the inferred surface dipole field $B_{\rm dip, surf}$ obtained from Eq.~\ref{Eq: Bdip}.
    \item For the $\epsilon_{\rm outer}$ heating profile we adopt $R_{\rm dyn} = R_{\rm heat} = r(P = 1 \text{ Mbar})$. These points are marked by filled circles in Fig.~\ref{Fig: HJ profiles}. This ensures that all inflation heating occurs outside the dynamo region (Eq.~\ref{Eq: B scaling law}) and also maximizes the distance between $R_{\rm dyn}$ and $R_{\rm P}$, yielding the smallest possible $B_{\rm dip, surf}$.
\end{enumerate}
These definitions are not intended to imply that the two heating prescriptions have intrinsically different physical dynamo boundaries. Rather, they aim to capture the limiting effects of the assumed dynamo boundary and thereby maximize or minimize the inferred magnetic field strength.

A commonly used alternative definition is $R_{\rm dyn} = R_{\rm P}(1-0.17/M_{\rm P})$, where $M_{\rm P}$ is the planet mass in Jupiter masses \citep[e.g.,][]{reiners_magnetic_2010, yadav_estimating_2017, hori_linkage_2021}. These radii are marked by triangles in Fig.~\ref{Fig: HJ profiles}. This prescription assumes that the dynamo radius is set by the pressure of hydrogen metallization, but is parameterized solely by the planetary mass and calibrated to Jupiter, with the dynamo extending to $\sim0.83\,\mathrm{R_J}$. In other words, it neglects the alkali ionization happening in HJs. As a consequence, for inflated HJs ($\epsilon \neq 0$) we generally have
\begin{equation*}
r(P=1\,{\rm Mbar}) \lesssim R_{\rm P}\left(1-\frac{0.17}{M_{\rm P}}\right) < r(\mathrm{Rm}=50),
\end{equation*} 
showing that this prescription lies between our two limiting definitions. For cold planets and non-inflated HJs, $r(P=1\,{\rm Mbar}) \sim R_{\rm P}\left(1-0.17/M_{\rm P}\right)$.

Our previous work \citep{elias-lopez_rossby_2025} adopted only $R_{\rm dyn}=r(P=1~{\rm Mbar})$. Using $r(\mathrm{Rm}=50)$ instead produces similar values of $B_{\rm rms, dyn}$, slightly smaller because the dynamo volume is larger, but systematically larger surface dipole fields. The increase ranges from a factor of $\sim1.1$ for compact planets ($R_{\rm P}\lesssim1.1\,\mathrm{R_J}$) up to $\sim6$ for highly inflated ones ($R_{\rm P}\gtrsim1.8\,\mathrm{R_J}$), illustrating the sensitivity of the inferred surface field to the adopted dynamo boundary.

Ideally, dynamo simulations spanning the metallic hydrogen region and the weakly conducting molecular envelope would be required to quantify the radial distribution of electric currents. Anelastic simulations with radially varying conductivity show that dynamo action is concentrated in the highly conducting interior, while weaker currents and secondary induction persist through the conductivity transition \citep{duarte_anelastic_2013, gastine_explaining_2014, dietrich_anelastic_2018, wicht_dynamo_2019}. The transition to an effectively current-free region is therefore gradual, making the dynamo surface particularly ambiguous for HJs, where it may extend over a substantial fraction of the planetary radius. For this work, a practical definition of $R_{\rm dyn}$ is required to assume a potential field beyond it and extrapolate the dynamo-generated field to the surface dipole field through Eq.~\ref{Eq: Bdip}.

\subsection{Simulation grid and interpolator}\label{Sec: Simulation grid and interpolator}

To construct the simulation grid, we used the planetary mass, $M_{\rm P}$, received stellar flux, $F_{\rm irr}$, and heating efficiency, $\epsilon$, as independent variables. We adopted logarithmically spaced values spanning $0.3 \leq M_{\rm P}(\mathrm{M_J}) \leq 12$, $10^7 \leq F_{\rm irr} ({\rm erg~ s}^{-1} {\rm cm}^{-2}) \leq 1.6\times10^{10}$, and $0 \leq \epsilon(\%) \leq 100$. The resulting $(F_{\rm irr}, M_{\rm P}, \epsilon)$ grid is shown in the left panel of Fig.~\ref{Fig: Simulation grid and method}. At sufficiently high combinations of $F_{\rm irr}$ and $\epsilon$, the models undergo runaway mass loss and completely evaporate \citep{batygin_evolution_2011}, which we exclude from the interpolation grid. The location of this evaporation boundary depends strongly on the planetary mass, as seen in Fig.~\ref{Fig: Simulation grid and method}; the low-mass planet slices contain larger regions without grid points. We computed separate grids for the $\epsilon_{\rm inner}$ and $\epsilon_{\rm outer}$ heating prescriptions described in Sect.~\ref{Sec: Evolutionary model}, each containing approximately 5000 evolution models. For every model we stored the planetary radius, $R_{\rm P}$, together with $B_{\rm rms,dyn}$ (Eq.~\ref{Eq: B scaling law}) and $B_{\rm dip,surf}$ (Eq.~\ref{Eq: Bdip}) for orbital periods between 0.1 and 100 days.

The interpolation grid is based on the evolved models at 5 Gyr. Inflated planets typically reach a quasi-steady state after $\sim$10--100 Myr, after which their radius and internal structure, including the convective velocity profile, evolve negligibly, making the predicted magnetic fields effectively age independent. This follows from the model assumptions that both the heating (i.e., $\epsilon$ and $q_{\rm heat}(r)$) and the incident stellar irradiation, $F_{\rm irr}$, remain approximately constant throughout the evolution. Young planets or systems with time-dependent heating or irradiation histories may therefore deviate from these predictions (see App.~\ref{App: Interior model caveats} for a more detailed discussion).

We then constructed linear interpolators, $\mathcal{I}_q$, for each quantity $q$ using {\tt SciPy v1.16} \citep{virtanen_scipy_2020}. Since the heating efficiency is not directly observable, the interpolators use the observed planetary radius, mass, received stellar flux, and orbital period ($R_{\rm P, obs}$, $M_{\rm P, obs}$, $F_{\rm irr, obs}$, and $P_{\rm orb, obs}$) as inputs, where $F_{\rm irr, obs}$ is derived from the stellar radius, effective temperature, and orbital separation. The right panel of Fig.~\ref{Fig: Simulation grid and method} illustrates the procedure. For each heating prescription, the interpolators return $\epsilon$, $B_{\rm rms,dyn}$, and $B_{\rm dip,surf}$. The NASA Exoplanet Archive \citep{christiansen_nasa_2025} contains 620 exoplanets with measured $M_{\rm P,obs}$, $R_{\rm P,obs}$, $F_{\rm irr,obs}$, and $P_{\rm orb,obs}$ that lie within our grid boundaries, including $0.1 \leq P_{\rm orb, obs}(\mathrm{days}) \leq 100$. Interpolating these observed parameters yields estimates of $\epsilon$, $B_{\rm rms,dyn}$, and $B_{\rm dip,surf}$ for each planet.

\section{Results}\label{Sec: Results}

To evaluate how much inflation is required for each HJ, we interpolate one further quantity: the non-inflated radius with $\epsilon=0$, denoted by $R_{\epsilon=0}$. This corresponds to the evolved radius at 5 Gyr of a planet with a given $M_{\rm P}$ and $F_{\rm irr}$ in the absence of anomalous internal heating. The choice of 5 Gyr is intended to be representative of the bulk of the observed HJ population and, within our modeling framework, yields nearly age-independent predictions for sufficiently inflated planets because the deposited heat balances the intrinsic cooling, producing a quasi-steady radius after $\sim0.1$--1 Gyr. This approximation relies on the assumptions that both the heating efficiency and the incident stellar irradiation remain approximately constant throughout the evolution. Unlike sufficiently heated models, however, planets with $\epsilon=0$ continue to cool and contract with age; the resulting age dependence of $R_{\epsilon=0}$ and its implications for our adopted reference radius are discussed in App.~\ref{App: Non-inflated models and threshold choice}. In terms of the interpolation grid, this corresponds to the $z=0$ plane in Fig.~\ref{Fig: Simulation grid and method}.

A fraction of the observed HJ population has radii smaller than $R_{\epsilon=0}$ and therefore cannot be interpolated. These planets are not necessarily inconsistent with the inflation scenario adopted here, but may instead reflect physical ingredients not included in our evolutionary models, such as larger $Z$, different atmospheric opacities, a different EoS, or past mass loss, particularly important for planets with $M_{\rm P}\lesssim0.5\,\mathrm{M_J}$ \citep[e.g.,][]{burrows_possible_2007, thorngren_mass-metallicity_2016, sestovic_investigating_2018}. In particular, our models adopt a relatively low envelope metallicity, whereas many giant exoplanets are expected to be substantially more metal-enriched, naturally producing smaller radii \citep{thorngren_mass-metallicity_2016, muller_mass-radius_2024}. Moreover, the discarded sample is dominated by weakly irradiated planets ($F_{\rm irr}<5\times10^8\,\mathrm{erg\,cm^{-2}\,s^{-1}}$), for which little or no inflation is expected, together with a population of massive HJs that remain compact despite their irradiation. Because this work aims to investigate how radius inflation modifies the magnetic field relative to the non-inflated case, we excluded these planets, leaving approximately 426 HJs with magnetic field estimates.

\begin{figure}[t]
\centering
\includegraphics[width=\hsize]{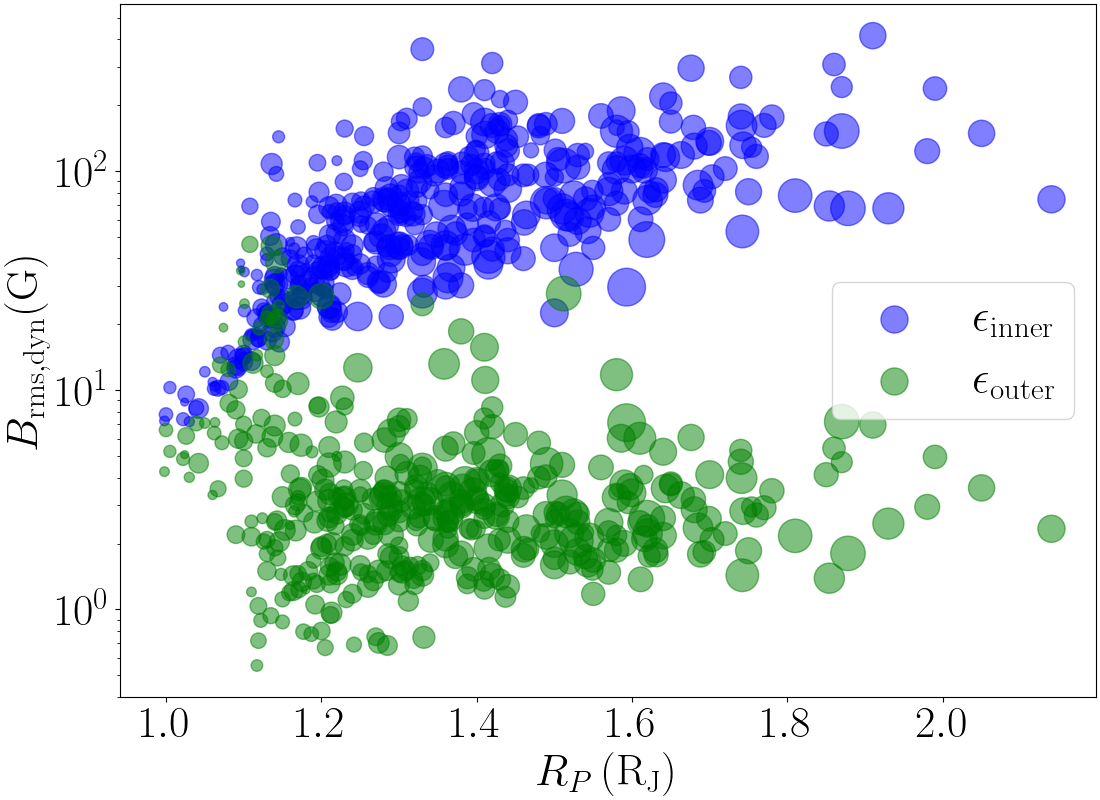}
\includegraphics[width=\hsize]{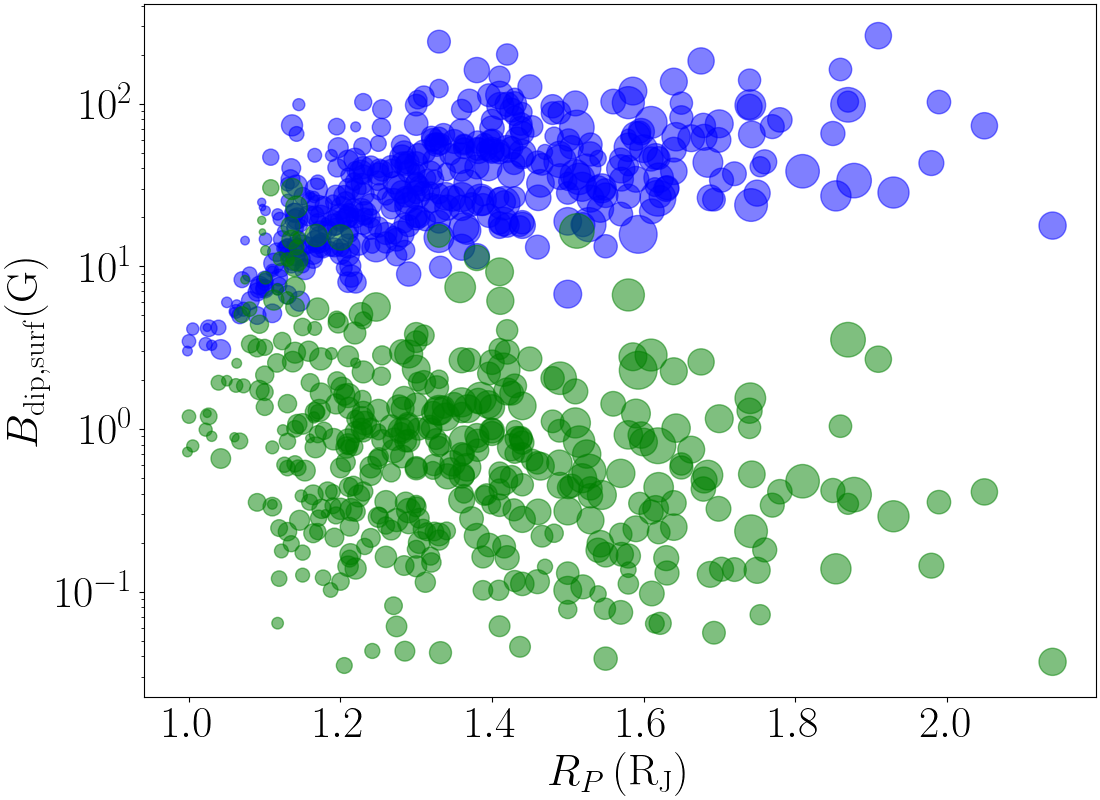}
\caption{Interpolated magnetic field estimates at the dynamo radius (top) and planetary surface (bottom), for the HJ population, as a function of $R_{\rm P}$. The size of the markers scales with $(F_{\rm irr})^{1/2}$ (with $F_{\rm irr}$ ranging from 1.1$\times10^7$ to 1.4$\times10^{10}$ erg cm$^{-2}$ s$^{-1}$). Blue and green markers show the $\epsilon_{\rm inner}$ and $\epsilon_{\rm outer}$ estimates, respectively.}
\label{Fig: Magnetic field HJ interpolation}
\end{figure} 

In Fig.~\ref{Fig: Magnetic field HJ interpolation} we show the interpolated values of $B_{\rm rms,dyn}$ and $B_{\rm dip,surf}$. The two heating prescriptions produce distinct magnetic-field populations. For the $\epsilon_{\rm inner}$ models, our predictions agree with \citet{yadav_estimating_2017}: most inflated HJs have $B_{\rm rms,dyn}\sim10^1$--$10^3$ G, typically around $10^2$ G, and $B_{\rm dip,surf}\sim10^1$--$10^2$ G, significantly stronger than Jupiter's $4.2$ G surface dipole field. In contrast, the $\epsilon_{\rm outer}$ models yield $B_{\rm rms,dyn}\sim10^0$--$10^2$ G and $B_{\rm dip,surf}\sim10^{-1}$--$10^1$ G, typically one to two orders of magnitude weaker than the $\epsilon_{\rm inner}$ predictions and comparable to or weaker than Jupiter's field.

Compared with \citet{yadav_estimating_2017}, our $\epsilon_{\rm inner}$ predictions exhibit a much stronger dependence on planetary radius because the interpolated heating efficiency is the value required to reproduce the observed radius, $R_{\rm P,obs}$. Larger radii therefore imply larger $\epsilon_{\rm inner}$ and stronger magnetic fields. In contrast, the $\epsilon_{\rm outer}$ models show a much weaker dependence on planetary radius, with at most a slight negative correlation. This dependence is absent in \citet{yadav_estimating_2017}, who adopted the irradiation-dependent heating efficiency of \citet{thorngren_bayesian_2018} (see App.~\ref{App: Th18 heating efficiency}), which depends only on $F_{\rm irr}$. A second difference arises from extrapolating the magnetic field from the dynamo region to the planetary surface. They used the common choices of $K_{\rm dip}=1/(2\sqrt{2})$ and $R_{\rm dyn}=(1-0.17/M_{\rm P})R_{\rm P}$, whereas we instead adopt the limiting prescriptions described in Sect.~\ref{Sec: Dynamo radius definition}.

\begin{figure*}[ht!]
\centering
\includegraphics[width=.95\hsize]{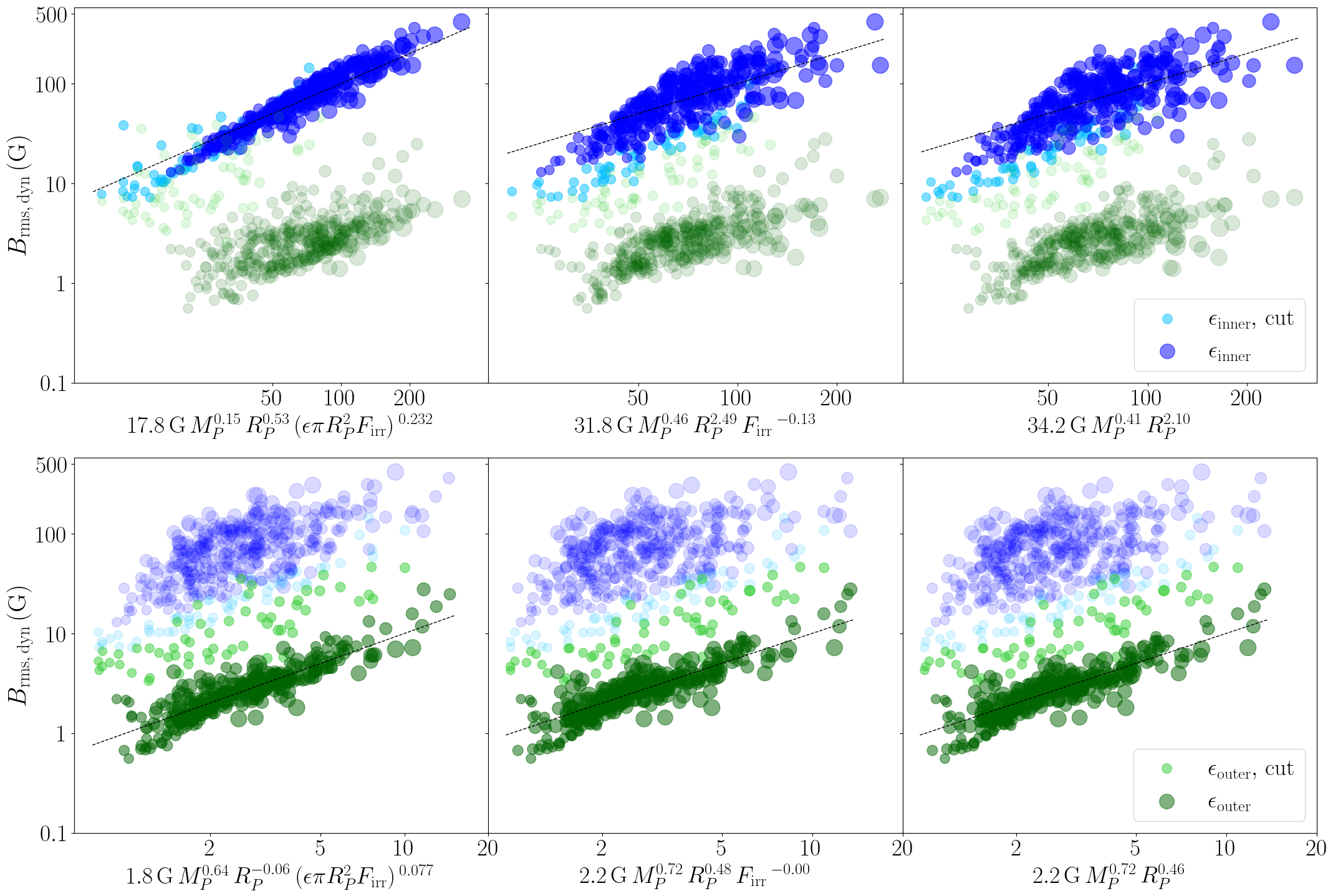}
\caption{Scaling-law fits for the predicted magnetic field strengths. The top panels show the $\epsilon_{\rm inner}$ (strong-field) models, while the bottom panels correspond to the $\epsilon_{\rm outer}$ (weak-field) models. The lighter colored points indicate planets with radii within 5\% of the non-inflated models ($R_{\rm obs} \leq 1.05\,R_{\epsilon=0}$), which are not included in the fits. The dashed lines show $x=y$. In the x-axis labels, $M_{\rm P}$ and $R_{\rm P}$ are in Jupiter units, $\epsilon\pi R_{\rm P}^2F_{\rm irr}$ in units of $\mathrm{L_J}$, and $F_{\rm irr}$ in units of $10^9$ erg cm$^{-2}$ s$^{-1}$. The fitted coefficients and their uncertainties are given in Table~\ref{Tab: Scaling laws}.}
\label{Fig: Fits comparisons}
\end{figure*}

\subsection{Scaling laws}\label{Sec: Scaling laws}

It is often practical to express model results as simple power-law scaling relations. We therefore fitted the interpolated values of $B_{\rm rms,dyn}$ and $B_{\rm dip,surf}$ as functions of $M_{\rm P}$, $R_{\rm P}$, $F_{\rm irr}$, and $L_{\rm P}$. To calibrate the fits, we excluded 11 planets requiring extreme heating efficiencies ($\epsilon>10\%$) and planets with only weak inflation ($R_{\rm P,obs}<1.05\,R_{\epsilon=0}$). This threshold is not intended to represent a physical boundary between inflated and non-inflated planets, since standard irradiated evolutionary models predict non-inflated radii up to $\sim1.3\,\mathrm{R_J}$ depending on the adopted atmospheric opacity, heavy-element abundance, and equation of state \citep{guillot_giant_1996, guillot_evolution_2002, arras_thermal_2006, burrows_possible_2007, fortney_planetary_2007}. Instead, it reflects our interpolation procedure, in which any radius excess above $R_{\epsilon=0}$ is attributed to anomalous internal heating. Similarly, heating efficiencies above $10\%$ lie well beyond the range typically inferred for the observed HJ population \citep{thorngren_bayesian_2018, sarkis_evidence_2021} and would require an implausibly efficient conversion of incident stellar irradiation into internal heat \citep{koll_atmospheric_2018}. The motivation and impact of these cuts are further discussed in App.~\ref{App: Non-inflated models and threshold choice}. After applying them, the final sample contains 356 planets.

Figure~\ref{Fig: Fits comparisons} compares the different fits for $B_{\rm rms,dyn}$, while Table~\ref{Tab: Scaling laws} lists the fitted coefficients for both $B_{\rm rms,dyn}$ and $B_{\rm dip,surf}$. The top panels correspond to the $\epsilon_{\rm inner}$ models and the bottom panels to $\epsilon_{\rm outer}$. From left to right, the number of fitted variables decreases, leading to larger scatter and a less random residual distribution. Although the analytical relations reproduce the overall behavior of the interpolated models, systematic residuals remain, particularly for $\epsilon_{\rm inner}$, indicating that a single power law cannot fully describe the magnetic-field dependence across the explored parameter space. We quantify the fit quality using the Pearson correlation coefficient $r$.\footnote{We define the Pearson $r = \frac{\sum_{i=1}^{n} (X_i - \bar{X})(Y_i - \bar{Y})}{\sqrt{\sum_{i=1}^{n} (X_i - \bar{X})^2} \sqrt{\sum_{i=1}^{n} (Y_i - \bar{Y})^2}}$, where $X$ and $Y$ correspond to the logarithms of interpolated and fitted magnetic field estimates, respectively.} The quoted coefficient uncertainties were obtained from the Monte Carlo procedure described in App.~\ref{App: Scaling laws error propagation} and therefore represent uncertainties in the fitted parameters rather than the intrinsic scatter of the relations.
\begin{table*}[h]

\centering
\caption{Coefficients for the different $B$ fits for the inflated HJ estimates with $R_{\rm obs} > 1.05\,R_{\epsilon=0}$. The $\epsilon=0$ fits use all 415 HJs with estimates and $\epsilon<10\%$.}
\begin{tabular}{ccccccc}
\hline \hline \\[-2.0ex]
\multicolumn{2}{c}{MRL fit (Eq.~\ref{Eq: Scaling law MRL})} 
    & $K$ (G) & $\alpha$ & $\beta$ & $\gamma$ & Pearson $r$ \\ \hline
\bluerow \multicolumn{7}{c}{} \\[-2ex]
\bluerow $\epsilon_{\rm inner}$ 
    & $B_{\rm rms,dyn}$  & 17.8$_{-2.6}^{+3.1}$ & 0.15$_{-0.09}^{+0.09}$ & 0.53$_{-0.49}^{+0.50}$ & 0.232$_{-0.039}^{+0.040}$ & 0.9580 \\ [.5ex]
\bluerow 
    & $B_{\rm dip,surf}$ & 9.9$_{-1.5}^{+1.7}$ & 0.38$_{-0.10}^{+0.09}$ & 0.11$_{-0.51}^{+0.50}$ & 0.236$_{-0.039}^{+0.041}$ & 0.9558 \\ [.5ex] \hline
\greenrow \multicolumn{7}{c}{} \\[-2ex]
\greenrow $\epsilon_{\rm outer}$ 
    & $B_{\rm rms,dyn}$  & 1.78$_{-0.31}^{+0.37}$ & 0.64$_{-0.10}^{+0.11}$ & -0.06$_{-0.51}^{+0.52}$ & 0.077$_{-0.043}^{+0.042}$ & 0.8888 \\ [.5ex]
\greenrow 
    & $B_{\rm dip,surf}$ & 0.73$_{-0.15}^{+0.20}$ & 1.49$_{-0.14}^{+0.14}$ & -1.67$_{-0.72}^{+0.72}$ & 0.045$_{-0.056}^{+0.056}$ & 0.9613 \\ [.5ex]

\hline \\[-2.0ex]
\multicolumn{2}{c}{MRF fit (Eq.~\ref{Eq: Scaling law MRF})} 
    & $K$ (G) & $\alpha$ & $\beta$ & $\gamma$ & Pearson $r$ \\ \hline
\bluerow \multicolumn{7}{c}{} \\[-2ex]
\bluerow $\epsilon_{\rm inner}$ 
    & $B_{\rm rms,dyn}$  & 31.8$_{-4.3}^{+5.0}$ & 0.461$_{-0.086}^{+0.084}$ & 2.49$_{-0.49}^{+0.48}$ & -0.132$_{-0.072}^{+0.073}$ & 0.8113 \\ [.5ex]
\bluerow 
    & $B_{\rm dip,surf}$ & 17.4$_{-2.3}^{+2.7}$ & 0.703$_{-0.089}^{+0.083}$ & 2.21$_{-0.48}^{+0.47}$ & -0.160$_{-0.074}^{+0.071}$ & 0.8282 \\ [.5ex] \hline
\greenrow \multicolumn{7}{c}{} \\[-2ex]
\greenrow $\epsilon_{\rm outer}$ 
    & $B_{\rm rms,dyn}$  & 2.23$_{-0.31}^{+0.36}$ & 0.719$_{-0.091}^{+0.090}$ & 0.48$_{-0.48}^{+0.46}$ & -0.004$_{-0.078}^{+0.078}$ & 0.8776 \\ [.5ex]
\greenrow 
    & $B_{\rm dip,surf}$ & 0.80$_{-0.15}^{+0.19}$ & 1.52$_{-0.12}^{+0.12}$ & -1.18$_{-0.69}^{+0.68}$ & -0.023$_{-0.10}^{+0.10}$ & 0.9596 \\ [.5ex]

\hline \\[-2.0ex]
\multicolumn{2}{c}{MR fit (Eq.~\ref{Eq: Scaling law MR})} 
    & $K$ (G) & $\alpha$ & $\beta$ & Pearson $r$ & \\ \hline
\bluerow \multicolumn{7}{c}{} \\[-2ex]
\bluerow $\epsilon_{\rm inner}$ 
    & $B_{\rm rms,dyn}$  & 34.2$_{-4.1}^{+4.7}$ & 0.410$_{-0.078}^{+0.078}$ & 2.10$_{-0.38}^{+0.37}$ & 0.7592 & \\ [.5ex]
\bluerow 
    & $B_{\rm dip,surf}$ & 19.0$_{-2.2}^{+2.7}$ & 0.638$_{-0.080}^{+0.077}$ & 1.73$_{-0.38}^{+0.38}$ & 0.7811 & \\ [.5ex] \hline
\greenrow \multicolumn{7}{c}{} \\[-2ex]
\greenrow $\epsilon_{\rm outer}$ 
    & $B_{\rm rms,dyn}$  & 2.24$_{-0.30}^{+0.33}$ & 0.716$_{-0.089}^{+0.086}$ & 0.46$_{-0.39}^{+0.39}$ & 0.8772 & \\ [.5ex]
\greenrow 
    & $B_{\rm dip,surf}$ & 0.81$_{-0.14}^{+0.17}$ & 1.51$_{-0.12}^{+0.12}$ & -1.24$_{-0.54}^{+0.56}$ & 0.9595 & \\ [.5ex] \hline
\grayrow \multicolumn{7}{c}{} \\[-2ex]
\grayrow $\epsilon=0$ 
    & $B_{\rm rms,dyn}$  & 11.76$_{-0.82}^{+0.86}$ & 0.718$_{-0.054}^{+0.054}$ & 0.08$_{-0.23}^{+0.22}$ & 0.9812 & \\ [.5ex]
\grayrow 
    & $B_{\rm dip,surf}$ & 6.55$_{-0.46}^{+0.49}$ & 0.889$_{-0.058}^{+0.057}$ & -0.06$_{-0.23}^{+0.23}$ & 0.9865 & \\ [.5ex]

\hline \hline \\[-2.0ex]
\end{tabular}
\label{Tab: Scaling laws}
\end{table*}

\subsubsection{The MRL relation}\label{Sec: MRL scaling law}

Inspired by \citet{reiners_surprisingly_2009} and \citet{yadav_estimating_2017}, we first fitted
\begin{equation}
    B [\text{G}] = K \, \left( \cfrac{M_{\rm P}}{\mathrm{M_J}} \right)^\alpha \,  \left( \cfrac{R_{\rm P}}{\mathrm{R_J}} \right)^\beta\, \left( \cfrac{\epsilon \pi R_P^2F_{\rm irr}}{\mathrm{L_J}}\right)^{\gamma},
    \label{Eq: Scaling law MRL}
\end{equation}
which we denote as the MRL fit (i.e., mass-radius-luminosity fit), and we similarly use $L_{\rm P}=\epsilon \pi R_{\rm P}^{2}F_{\rm irr}$ as the injected heat. This parametrization allows a direct comparison with R09 (Eq.~\ref{Eq: Bdyn scaling law Reiners10}), which has $\alpha=1/6$, $\beta=-7/6$ and $\gamma=1/3$.

For the $\epsilon_{\rm inner}$ models, the MRL relation reproduces the interpolated values well ($r\simeq0.96$). The fitted heating exponent ($\gamma\simeq0.23$) is smaller than the R09 value of $1/3$, indicating a slightly weaker dependence on the deposited heating power. The mass exponent ($\alpha\simeq0.15$) is close to the R09 value of $1/6$, while the largest difference is the positive radius exponent ($\beta\simeq0.53$ instead of $-7/6$), reflecting the correlation between radius inflation and heating efficiency in the $\epsilon_{\rm inner}$ models, as discussed above.

For the $\epsilon_{\rm outer}$ models, $\beta$ and $\gamma$ are statistically consistent with zero for $B_{\rm rms,dyn}$ and $B_{\rm dip, surf}$, respectively, indicating that the magnetic field depends primarily on either the radius or the injected heat. This reflects the fact that heat deposited above the metallic hydrogen region contributes little to the deep convective luminosity that powers the dynamo \citep[e.g.,][]{elias-lopez_planetary_2025, vigano_inflated_2025}. These differences from R09 are expected because the latter does not account for the structural modifications induced by intense stellar irradiation and heat deposition. In inflated HJs, the planetary radius traces the interior thermal state and therefore becomes a much stronger predictor of the magnetic field than in non-inflated giant planets.

\subsubsection{Scaling with observables}\label{Sec: MRF scaling law}

For observational applications, it is desirable to use only directly measurable quantities. We therefore replaced the deposited heat by the incident stellar flux,
\begin{equation}
    B [\text{G}] = K  \, \left( \cfrac{M_{\rm P}}{\mathrm{M_J}} \right)^\alpha \,  \left( \cfrac{R_{\rm P}}{\mathrm{R_J}} \right)^\beta \, \left( \cfrac{F_{\rm irr}}{10^9 \, \mathrm{erg \, cm^{-2} \, s^{-1}}}\right)^{\gamma},
    \label{Eq: Scaling law MRF}
\end{equation}
which we denote as the MRF fit (i.e., mass-radius-flux fit). The fits can be seen in Fig.~\ref{Fig: Fits comparisons} and the coefficients are listed in Table~\ref{Tab: Scaling laws}. $\gamma$ is small: it is consistent with zero for $\epsilon_{\rm outer}$ and only slightly negative for $\epsilon_{\rm inner}$, indicating that once the planetary mass, radius, and heating prescription are specified, the incident stellar flux adds little predictive power. Physically, this suggests that the effect of irradiation is already largely encoded in the inflated radius.

For the $\epsilon_{\rm inner}$ models, replacing the deposited heating with $F_{\rm irr}$ reduces the fit quality from $r\simeq0.96$ to $r\simeq0.81$, showing clear systematic residuals. This implies that $F_{\rm irr}$ cannot capture well the information contained in the deposited heating. This agrees with \citet{thorngren_bayesian_2018}, who found that the heating efficiency cannot be described by a simple power law of $F_{\rm irr}$. Although they proposed a Gaussian prescription that better reproduces the average trend, it still exhibits a large intrinsic scatter, implying that planets receiving similar irradiation may require substantially different heating efficiencies to reproduce their observed radii. In contrast, the $\epsilon_{\rm outer}$ fits deteriorate only slightly, consistent with their weaker dependence on the deposited heating.

\subsubsection{Scaling with only mass and radius}\label{Sec: MR scaling law}

Motivated by these results, we finally considered a simpler relation involving only the planetary mass and radius,
\begin{equation}
    B[\text{G}] = K \, \left( \cfrac{M_{\rm P}}{\mathrm{M_J}} \right)^\alpha \, \,  \left( \cfrac{R_{\rm P}}{\mathrm{R_J}} \right)^\beta
    \label{Eq: Scaling law MR}
\end{equation}
which we denote as the MR fit (i.e., mass-radius fit). The resulting fits are comparable to the MRF relations, with only a modest decrease in the Pearson coefficients despite the removal of one free parameter. This indicates that, once the deposited heating is removed from the parametrization, including $F_{\rm irr}$ provides only a limited improvement. Nevertheless, both the MRF and MR relations exhibit substantially larger scatter than the MRL fits, particularly for the $\epsilon_{\rm inner}$ models, as can be appreciated in Fig.~\ref{Fig: Fits comparisons}. The simplified scaling laws therefore reproduce the global trends in the magnetic field, but they do not capture the full planet-to-planet variation present in the interpolated models. 

Despite this limitation, the MR relation is particularly attractive because it depends only on planetary mass and radius, two quantities commonly measured for transiting HJs. It therefore provides a convenient first-order estimate of $B_{\rm rms,dyn}$ and $B_{\rm dip, surf}$, while the interpolation framework should be preferred whenever greater accuracy is required. App.~\ref{App: Specific exoplanet comparison} compares the MR, interpolated, and R09 magnetic-field estimates for three representative HJs.

\section{Discussion}\label{Sec: Discussion}

The magnetic field estimates presented here inherit uncertainties from the underlying planetary interior models, the dynamo scaling laws, and the observational data. A detailed analysis of these effects is presented in App.~\ref{App: Interior model caveats}. Concerning the interior models, we investigated the effects of the age used to construct the grid, equation of state, and heavy-element content in the envelope. In brief, we found that the predicted magnetic field strengths are relatively insensitive to the assumed age of inflated HJs and to the choice of modern H-He equation of state, whereas the inferred metallicity can have a larger indirect impact through its degeneracy with the heating efficiency required to reproduce the observed planetary radius. We emphasize that the present work does not aim to evaluate the role of the core mass, the heavy-element distribution, or the value of $\Sigma_{\rm irr}$, whose effects have been investigated elsewhere \citep{thorngren_mass-metallicity_2016,vigano_inflated_2025}. Another possible neglected influence is compositional gradients, which might lead to stably stratified layers in the form of helium rain or \textit{fuzzy cores}. These parameters are expected to contribute to the observed diversity of HJs and, consequently, to the scatter around the proposed scaling laws.

Additional uncertainty arises from the applicability of the dynamo scaling law itself. HJs may operate in a dynamo regime that differs significantly from those of Jupiter and the low-mass objects used to calibrate the existing relations. In particular, the metallic hydrogen region is surrounded by a partially ionized envelope, where thermal ionization of alkali metals maintains a non-negligible electrical conductivity \citep[][see also Sect.~\ref{Sec: Dynamo radius definition}]{kumar_ionization_2021, dietrich_magnetic_2022}. This smooth conductivity transition allows electrical currents to extend to lower pressures than in the Solar System giants, making the definition of dynamo radius more complex (as argued in Sect.~\ref{Sec: Dynamo radius definition}). Furthermore, atmospheric induction driven by ionized winds can interact with the deep magnetic field and deposit Ohmic heat into the envelope \citep{perna_ohmic_2010,batygin_inflating_2010,rogers_hottest_2017}. Together with the intense stellar irradiation, these effects produce a more complex internal luminosity profile than in isolated giant planets or low-mass stars, where the scaling laws have primarily been tested.

The reformulation of the original scaling law by \cite{reiners_surprisingly_2009} (Eq.~\ref{Eq: Bdyn scaling law Reiners10}) already assumes an intrinsic dispersion in $B_{\rm rms}$ of $\sim 60\%$. Overall, we similarly assume that the combined uncertainties associated with the interior models and the scaling laws are typically of a factor of 2, which we use to derive the uncertainties of the scaling law parameters (see App.~\ref{App: Interior model caveats} and \ref{App: Error estimation}). Therefore, the magnetic field estimates presented in this work should be regarded as order-of-magnitude predictions.

\begin{figure}[t]
\centering
\includegraphics[width=\hsize]{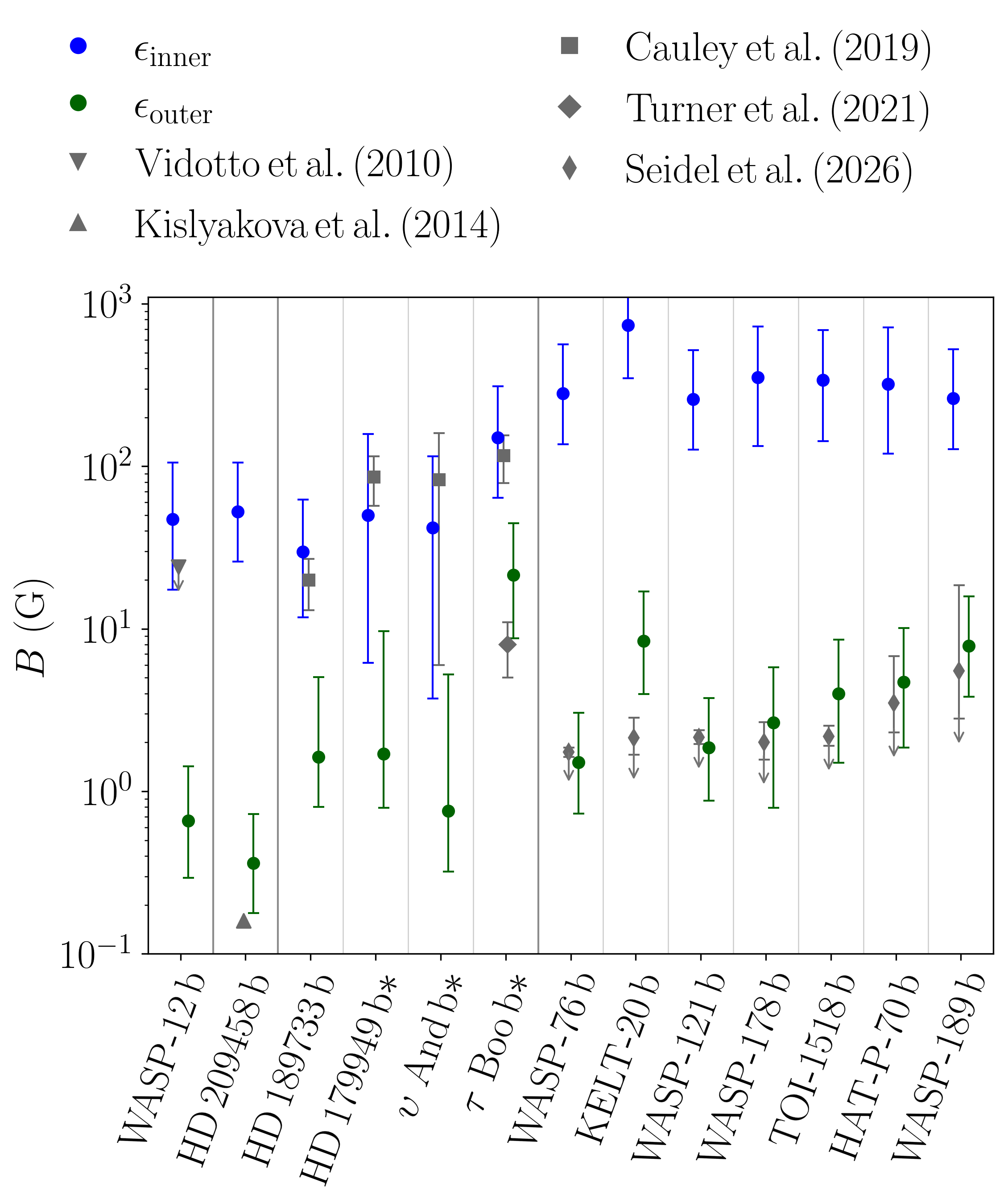}
\caption{Comparison of our interpolated magnetic field estimates and indirect magnetic field constraints reported in the literature. We show $2 B_{\rm dip, surf}$ when comparing with the \cite{vidotto_early_2010, cauley_magnetic_2019} and \cite{turner_search_2021} SPI estimates, and $10 B_{\rm dip, surf}$ when comparing with the \cite{seidel_magnetic_2026} atmospheric maximum estimations (see text). Both the exoplanetary parameters and the magnetic estimates are shown in App.~\ref{App: Comparison with literature values}.}
\label{Fig: B comparison with literature}
\end{figure} 

\subsection{Comparison with literature values}

To date, magnetic field estimates have been reported for a limited number of HJs using a variety of observational approaches, including chromospheric SPI signatures \citep{cauley_magnetic_2019} and, more recently, atmospheric magnetic drag inferred from high-resolution transmission spectroscopy \citep{seidel_magnetic_2026}. In Fig.~\ref{Fig: B comparison with literature}, we compare our estimates with the results of these two studies. Other observational approaches, including Ly-$\alpha$ escape modeling and radio searches, have provided other constraints for single planets \citep{vidotto_early_2010, kislyakova_magnetic_2014, turner_search_2021}.

The planets analyzed by \citet{cauley_magnetic_2019} that exhibit orbitally modulated Ca {\sc II} emission are HD~189733~b, HD~179949~b, $\tau$~Boo~b, and $\upsilon$~And~b. Assuming their nominal conversion efficiency of $E_{\rm CaIIK}/E_{\rm tot}=0.20\%$, their inferred polar magnetic fields span $\sim20$--120 G, broadly consistent with the estimates of \citet{yadav_estimating_2017} and with our $\epsilon_{\rm inner}$ models (noting that $B_{\rm pole,surf}=2B_{\rm dip,surf}$). However, this efficiency is poorly constrained and represents an important uncertainty in the SPI-derived fields. Of these systems, only HD~189733~b has a directly measured radius ($R_{\rm P}=1.14\,\mathrm{R_J}$), allowing a direct comparison with our models. The other three planets do not transit, and their radii must therefore be inferred from mass--radius relations \citep[e.g.,][]{chen_probabilistic_2016, muller_mass-radius_2024}, which have large uncertainties. For $\tau$~Boo~b, however, the inferred radius falls below the lower boundary of our interpolation grid, so only the analytical scaling laws can be applied. Conversely, for HD~179949~b, $\tau$~Boo~b, and $\upsilon$~And~b, we can estimate the degree of radius inflation that would be required for our models to reproduce the SPI-inferred magnetic fields. Such comparisons should nevertheless be interpreted with caution, since both the dynamo radius and convective heat flux, and hence the predicted magnetic field, are sensitive to the assumed planetary radius and inflation level.

A second comparison can be made with \citet{seidel_magnetic_2026}, who inferred upper limits on the atmospheric field strength, $B_{\rm atm, max}$, of approximately $2$--$5$ G for seven inflated ultra-HJs. These constraints are based on atmospheric wind velocities measured through transmission spectroscopy, assuming that sufficiently strong magnetic fields would brake the ionized flow \citep{menou_magnetic_2012}. For an order-of-magnitude comparison, we convert our $B_{\rm dip,surf}$ estimates to approximate maximum atmospheric fields by adopting $B_{\rm atm, max}/B_{\rm dip, surf}\sim10$ (for Jupiter, this ratio is $18/4.2\simeq4.3$). This factor is intended only as a rough conversion and should not be interpreted as a universal relation for HJs, as it depends on the magnetic-field geometry and atmospheric dynamics: the inferred atmospheric fields may include locally induced fields generated by ionized winds \citep{rogers_hottest_2017, dietrich_magnetic_2022, soriano-guerrero_non-ideal_2025}, so the contribution from the deep dynamo could be substantially weaker. With this approximate scaling, our weak-field scenario remains broadly consistent with the atmospheric constraints.

\subsection{How to apply these results}

The results presented above provide magnetic field estimates for giant exoplanets, particularly evolved HJs. The most accurate approach is to use our interpolator (see Code availability), which requires $M_{\rm P}$, $R_{\rm P}$, $F_{\rm irr}$, and $P_{\rm orb}$ as input. Alternatively, the scaling laws presented in Sect.~\ref{Sec: Scaling laws} may be used with the coefficients listed in Table~\ref{Tab: Scaling laws}. For the simplest estimate available, the MR fits are:
\begin{gather}
    B_{\rm dip, surf}^{\rm strong} = 19_{-10}^{+19} \left( \frac{M_{\rm P}}{\mathrm{M_J}} \right)^{0.64 \pm 0.08} \left( \frac{R_{\rm P}}{\mathrm{R_J}} \right)^{1.7\pm0.4} ~{\rm G}
    \label{Eq: Bdip scaling law inner}\\
    B_{\rm dip, surf}^{\rm weak} = 0.81_{-0.43}^{+0.83} \left( \frac{M_{\rm P}}{\mathrm{M_J}} \right)^{1.5 \pm 0.1} \left( \frac{R_{\rm P}}{\mathrm{R_J}} \right)^{-1.2 \pm 0.6} ~{\rm G}
    \label{Eq: Bdip scaling law outer}
\end{gather}
corresponding to the $\epsilon_{\rm inner}$ (strong-field) and $\epsilon_{\rm outer}$ (weak-field) scenarios, respectively. For inflated HJs, Eq.~\ref{Eq: Bdip scaling law inner} predicts surface dipole fields of several tens to hundreds of gauss, consistent with the largest inferred values from SPI-modulated activity \citep{cauley_magnetic_2019}, whereas Eq.~\ref{Eq: Bdip scaling law outer} yields fields typically below Jupiter's, comparable to the upper limits inferred spectroscopically by \cite{seidel_magnetic_2026}. For non-inflated gas giants (e.g., $R_{\rm P} < 1.1 \, \mathrm{R_J}$), we instead recommend:
\begin{equation}
    B_{\rm dip, surf}^{\rm \epsilon=0} = 6.5_{-3.3}^{+6.6} \left( \frac{M_{\rm P}}{\mathrm{M_J}} \right)^{0.89 \pm 0.06} \left( \frac{R_{\rm P}}{\mathrm{R_J}} \right)^{-0.1\pm0.2} ~{\rm G}
    \label{Eq: Bdip scaling law irr}
\end{equation}
which reproduces our unheated models at 5 Gyr, and is compatible with the Jovian equatorial dipole. To obtain the errors for the constants, we have taken into account the uncertainty of the coefficients $K$ shown in Table~\ref{Tab: Scaling laws} and added a factor of 2 systematic error in quadrature coming from the interior modeling and magnetic scaling law, as discussed above and in App.~\ref{App: Error estimation}.

\section{Conclusions}\label{Sec: Conclusions}

We computed one-dimensional evolutionary models of irradiated and internally inflated gas giants to obtain their interior structures and predict the dynamo-generated magnetic fields of HJs. We explored two limiting heating scenarios depending on whether the inflationary heat is deposited within or above the dynamo region \citep[similar to][]{elias-lopez_rossby_2025}. Heat deposited within the dynamo region ($\epsilon_{\rm inner}$) enhances the convective power available to the dynamo and produces strong magnetic fields, whereas heat deposited above it ($\epsilon_{\rm outer}$) bypasses the metallic hydrogen layer, reducing the convective heat flux and yielding substantially weaker fields. These two prescriptions therefore provide upper and lower estimates of the magnetic field strengths expected from inflationary heating.

Based on these models, we constructed interpolation grids that estimate the heating efficiency and magnetic field of HJs from their observed properties. We also provide scaling laws to estimate magnetic fields from readily measured quantities. The simplest relations we provide (Eqs.~\ref{Eq: Bdip scaling law inner}, \ref{Eq: Bdip scaling law outer}, and \ref{Eq: Bdip scaling law irr}) are only dependent on $M_{\rm P}$ and $R_{\rm P}$ (see Code availability). Given the uncertainties associated with the applicability of the dynamo scaling law to HJs, the dynamo radius, and the interior models (e.g., composition, equation of state, adiabatic structure, and simplified time-independent heating prescriptions), the predicted field strengths should be regarded as order-of-magnitude estimates.

To remain agnostic about the inflation mechanism, we adopted idealized step-function heating profiles. The shallow-heating prescription ($\epsilon_{\rm outer}$) is motivated by the leading inflation mechanisms proposed for HJs, including Ohmic dissipation and atmospheric circulation-driven heating \citep{showman_atmospheric_2002, showman_atmospheric_2009, li_circulation_2010, youdin_mechanical_2010, batygin_inflating_2010, batygin_evolution_2011, perna_ohmic_2010, wu_ohmic_2012}. These mechanisms deposit most of their energy above the metallic hydrogen region \citep{komacek_structure_2017, knierim_shallowness_2022, vigano_inflated_2025}, contributing little to the deep convective dynamo and naturally favoring the weak-field scenario. Likewise, explanations based on enhanced atmospheric opacities \citep{burrows_possible_2007} or inhibited convection \citep{chabrier_heat_2007} reduce the planetary cooling rate rather than supplying deep internal heating, leading to similarly weak convective dynamos. Alternatively, tidal dissipation could provide internal heating at greater depths \citep[see ][for a review]{fuller_tidal_2024}, represented here with the $\epsilon_{\rm inner}$ models, which would enhance the dynamo.

Although multiple inflation processes may operate simultaneously \citep{sarkis_evidence_2021}, recent work suggests that shallow heating, irrespective of its physical origin, dominates HJ radius inflation \citep{schmidt_hot_2026}, favoring the weak-field scenario. Nevertheless, medium-to-strong fields cannot be ruled out. If the inflation mechanism and the deep dynamo are coupled, time-dependent heating could produce long-term magnetic-field variability and potentially reconcile different observational diagnostics \citep{vigano_inflated_2025}.

Several caveats should be kept in mind. Although our inflated HJs remain in the rapidly rotating regime \citep{elias-lopez_rossby_2025}, the applicability of the scaling law of \citet{christensen_energy_2009} to HJ interiors remains untested. This scaling reproduces the magnetic-field strengths of Earth, Jupiter, young contracting stars, and rapidly rotating low-mass stars, and is also consistent with the recent measurement of a $\gtrsim1.25$ kG magnetic field in the young $\sim12\,\mathrm{M_J}$ planet $\beta$~Pictoris~b \citep{ceballos_discovery_2026}. We emphasize that, although applying this scaling is common in the exoplanet literature, it has not been directly validated for the inflated HJs considered here. We assume time-independent heating and describe only the large-scale field generated by the deep interior dynamo. Furthermore, we do not account for small-scale atmospheric dynamos or non-linear MHD effects that may operate in highly ionized HJ atmospheres \citep{rogers_hottest_2017, dietrich_magnetic_2022, soriano-guerrero_non-ideal_2025}, nor for time-dependent coupled interior--dynamo evolution. 

Finally, we also assume a dipole-dominated field, which may be less accurate for the thinner dynamo regions in the $\epsilon_{\rm outer}$ models, where multipolar dynamos could produce substantially weaker dipoles \citep{wulff_new_2026}. For a multipolar dynamo, the scaling laws of \citet{wulff_new_2026} give, at the same driving power, $B_{\rm rms,dyn}^{\rm multipolar} \approx 0.73 \, B_{\rm rms,dyn}^{\rm dipolar}$ and $B_{\rm dip,surf}^{\rm multipolar} \approx 0.096 \, B_{\rm dip,surf}^{\rm dipolar}$.\footnote{Ratios of the multipolar and dipolar prefactors of the rms field in the dynamo volume, $0.471/0.646$, and of the dipole at the dynamo surface, $0.015/0.157$ \citep[Table 2 of][]{wulff_new_2026}.} While the results presented here assume a dipole-dominated dynamo, our public code (see below) lets the user choose between dipolar and multipolar dynamos and provides the multipolar estimates for all the planets studied here.

At present, the available observational constraints appear to favor different limiting heating scenarios. The SPI-derived magnetic-field estimates of \citet{cauley_magnetic_2019} lie closer to the upper end of our predicted range, whereas the atmospheric constraints of \citet{seidel_magnetic_2026}, which limit the atmospheric field to at most a few gauss, are more consistent with the $\epsilon_{\rm outer}$ models. Both approaches, however, involve important caveats. The atmospheric constraints are based on measured wind velocities but require assumptions to translate them into a maximum atmospheric magnetic field. At the same time, the SPI estimates rely on converting the observed chromospheric modulation into a planetary magnetic-field strength, which is model-dependent and particularly sensitive to the poorly constrained interaction efficiency. Radio searches have so far yielded no confirmed detections of planetary emission, despite predictions that fields of order $\sim100$ G could produce signals detectable with current facilities. Although these non-detections may result from physical or observational effects, including self-absorption, beaming, and temporal variability, and therefore do not rule out strong magnetic fields, they currently provide no observational support for them.

The differences between \citet{cauley_magnetic_2019} and \citet{seidel_magnetic_2026} may partly reflect the different planetary populations probed: the latter specifically targets ultra-HJs, whereas the former considers less extremely irradiated HJs. If this apparent correspondence between the observations and our limiting heating scenarios is physical, it may arise from differences in the depth of heat deposition. Current observational constraints remain insufficient to distinguish conclusively between the two limiting models. Moreover, the complexity of planetary heating and interior evolution allows a broad range of intermediate scenarios, including evolving irradiation, non-uniform heat deposition, and coupling between heating and dynamo processes. Future systematic magnetic measurements across a range of $T_{\rm eq}$ and $M_{\rm P}$, together with improved understanding of their interior structures, could help determine whether magnetic-field strengths depend systematically on irradiation or other planetary properties, thereby constraining the dominant inflation mechanism and its connection to magnetic-field generation.

\section*{Code availability}\label{Sec: Code availability}

The general tool developed in this work is publicly available at
\href{https://github.com/albert-elias-lopez/hj_magnetic_field_interpolator}{github.com/albert-elias-lopez/ hj\_magnetic\_field\_interpolator}.

\begin{acknowledgements}
This publication is part of the project "Blowing in the wind: exoplanetary atmospheres being carried away by stellar winds" (with project number VI.C.232.041 of the research programme "NWO Talent Programme VICI 2023"), which is financed by the Dutch Research Council (NWO). DV is supported by the European Research Council (ERC) under the European Union’s Horizon 2020 research and innovation programme (ERC Starting Grant "IMAGINE" No. 948582), by ``ERDF A way of making Europe'', from the programme Unidad de Excelencia María de Maeztu CEX2020-001058-M and MaX-CSIC Excellence Award MaX4 SOMMA-ICE, and from the Generalitat de Catalunya/CERCA. The authors thank Yamila Miguel, Keren Duer-Millner, Paula Wulff, and Camilla Pezzotti for fruitful discussions and comments, and the anonymous referee for their constructive feedback. 
\end{acknowledgements}

\bibliographystyle{aa}
\bibliography{biblio}

\begin{appendix}

\nolinenumbers

\section{R09 scaling law}
\label{App: R09 scaling law}

Relying on the original work of \cite{christensen_energy_2009} who calibrated dynamo scaling laws with observed magnetic fields in solar system planets and low-mass fast-rotating stars, \cite{reiners_surprisingly_2009} expressed the mean strength of the magnetic field at the dynamo surface, $B_{\rm dyn}$, in terms of the mass $M_{\rm P}$, internal convective luminosity $L_{\rm P}$, and radius $R_{\rm P}$ of the substellar object \footnote{For the rescaling, we adopt $\mathrm{M_J}=1.898\times10^{27}$ kg, $\mathrm{M_\odot}=1.989\times10^{30}$ kg, $L_J=4.61\times10^{24}$ erg s$^{-1}$ (corresponding to 7.5 W m$^{-2}$; \citealt{li_less_2018}), $L_\odot=3.82\times10^{33}$ erg s$^{-1}$, and $\mathrm{R_\odot}=9.73\,\mathrm{R_J}$.}:
\begin{equation}
    B_{\rm rms, dyn}^{\rm R09} = 23_{-13}^{+15} \left( \frac{M_{\rm P}}{\mathrm{M_J}} \right)^{1/6} \left( \frac{R_{\rm P}}{\mathrm{R_J}} \right)^{-7/6} \left( \frac{L_{\rm P}}{\mathrm{L_J}} \right)^{1/3}~{\rm G}
    \label{Eq: Bdyn scaling law Reiners10}
\end{equation}
We refer to this expression as the R09 scaling law and use it to compare with our results.

Despite its simplicity and limitations, the R09 scaling law has been widely applied to substellar objects and exoplanets. Its application to HJs, however, is complicated by the unknown convective luminosity $L_{\rm P}$ that powers the dynamo. While evolutionary models can estimate $L_{\rm P}$ for cold gas giants and brown dwarfs, radius inflation prevents its determination for HJs without additional assumptions. See Appendix A of \citet{vigano_inflated_2025} for further discussion and caveats.

\begin{figure}[h]
\centering
\includegraphics[width=\hsize]{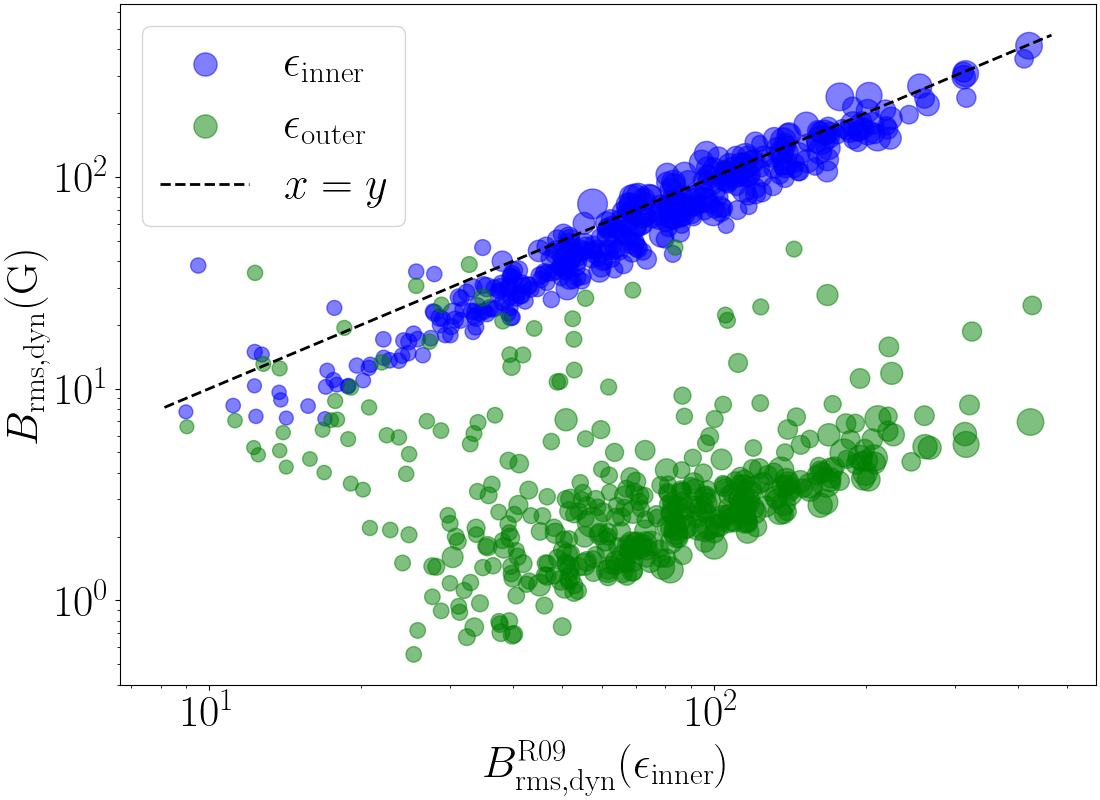}
\caption{Comparison between the $B_{\rm rms,dyn}$ values predicted by the \citet{reiners_surprisingly_2009} scaling law (x-axis) and the interpolated ones (y-axis) for each planet. For the R09 prediction, we assume that the internal luminosity is equal to the total injected heating power, $L_{\rm P}=\epsilon\pi R_{\rm P}^2F_{\rm irr}$, as in \citet{yadav_estimating_2017}. The dashed line indicates perfect agreement with the R09 prediction.}
\label{Fig: Comparison with reienrs10}
\end{figure} 

In Fig.~\ref{Fig: Comparison with reienrs10}, we compare our interpolated $B_{\rm rms,dyn}$ values with the prescription of \citet{yadav_estimating_2017}, who applied the R09 scaling law assuming that the convective luminosity available to the dynamo equals the injected heating power, $L_{\rm P}=\epsilon\pi R_{\rm P}^{2}F_{\rm irr}$. Since our $\epsilon_{\rm inner}$ models make the same assumption, the R09 prescription reproduces them reasonably well, with most planets lying close to the one-to-one relation. In contrast, the $\epsilon_{\rm outer}$ predictions are systematically weaker, reflecting the reduced coupling between the deposited heat and the deep convective region. Adopting the empirical heating efficiencies $\epsilon_{\rm Th18}$ (see App.~\ref{App: Th18 heating efficiency}) instead produces a similar overall trend but substantially larger scatter, because $\epsilon_{\rm Th18}$ depends only on $F_{\rm irr}$, whereas our interpolated efficiencies are constrained by the observed properties of each planet.

\section{Th18 heating efficiency}
\label{App: Th18 heating efficiency}

\citet{thorngren_bayesian_2018} derived an analytical relation for the heating efficiency as a function of incident stellar flux, $\epsilon(F_{\rm irr})$, that reproduces the population-level trend of HJ radii with irradiation (their Eq.~34):
\begin{equation}
    \epsilon_{\rm Th18}(\%) = 2.37^{+1.3}_{-0.26} \exp \left[ -\frac{\left(\log_{10} (F_{\rm irr}/F_0) - 0.14^{+0.060}_{-0.069}\right)^2}{2 \left( 0.37^{+0.038}_{-0.059}~\right)^2}~\right]~,
    \label{Eq: Th18 epsilon}
\end{equation}
where $F_0=10^9~{\rm erg~s^{-1}~cm^{-2}}$. This relation is non-monotonic, reaching a maximum efficiency of $\epsilon\simeq2.4\%$ at $T_{\rm eq}\approx1500$~K before declining to $\sim0.2\%$ at $T_{\rm eq}\approx2500$~K.

\begin{figure}[h]
\centering
\includegraphics[width=\hsize]{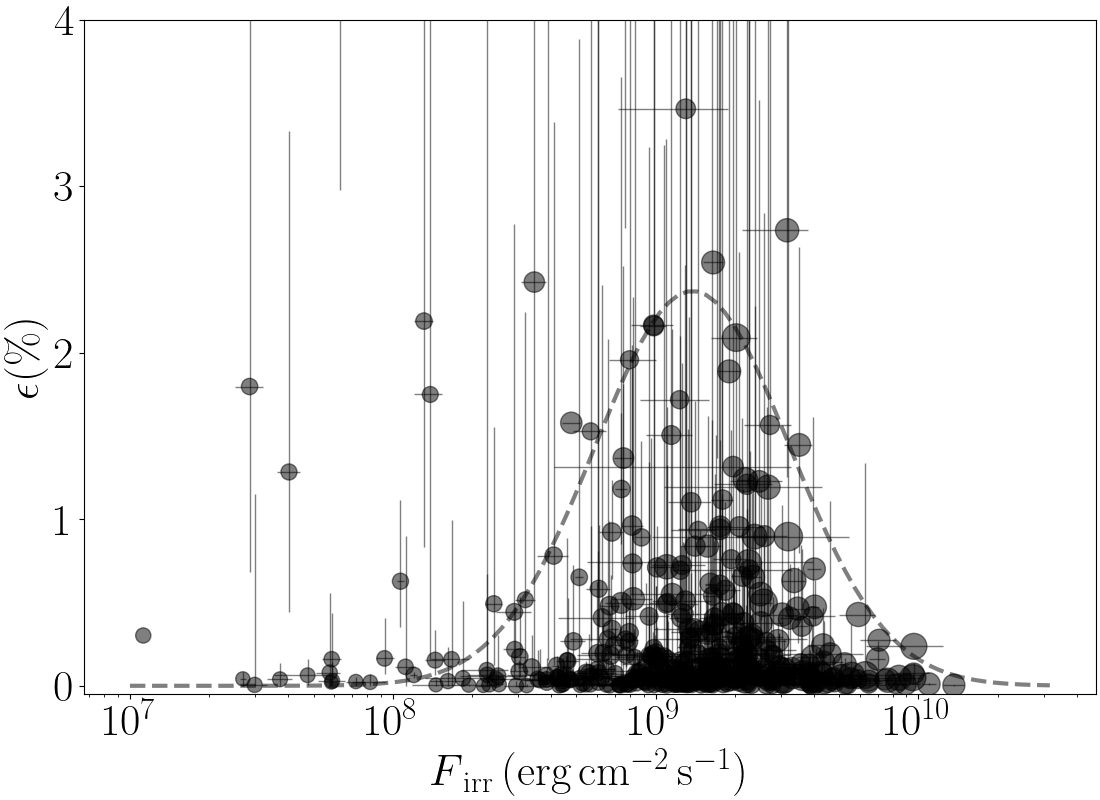}
\caption{Interpolated heating efficiencies for the internal heat deposition mechanism, $\epsilon_{\rm inner}$, as a function of incident stellar flux. Marker size scales with $R_{\rm P}^2$. The dashed line shows the $\epsilon_{\rm Th18}$ relation from \citet{thorngren_bayesian_2018} (Eq.~\ref{Eq: Th18 epsilon}).}
\label{Fig: Heating efficiency vs flux}
\end{figure}

Figure~\ref{Fig: Heating efficiency vs flux} shows our interpolated $\epsilon_{\rm inner}$ values as a function of $F_{\rm irr}$, together with the population-level relation of \citet{thorngren_bayesian_2018}. Despite the substantial planet-to-planet scatter, our inferred efficiencies recover the same general behavior found by \citet{thorngren_bayesian_2018} and \citet{sarkis_evidence_2021}: $\epsilon$ increases with irradiation, peaks around $(2$--$3)\times10^9~\mathrm{erg\,cm^{-2}\,s^{-1}}$, and decreases at higher fluxes. This turnover provides an important constraint on inflation mechanisms, since not all proposed mechanisms predict reduced heating efficiencies at high irradiation. The relatively large uncertainties in our individual estimates reflect the sensitivity of the interpolation to the observational uncertainties in $R_{\rm P}$, $M_{\rm P}$, and $F_{\rm irr}$. Our inferred efficiencies are nevertheless systematically lower than those of \citet{thorngren_bayesian_2018} and \citet{sarkis_evidence_2021}. This difference likely reflects the adopted planetary structure models. Their calculations used fully convective planets without distinct cores and allowed the heavy-element abundance to vary, whereas we adopt a $10\,M_\oplus$ inert core and a fixed envelope metallicity. As discussed in App.~\ref{App: Z dependence}, these structural and compositional assumptions affect the radius obtained for a given heating efficiency and therefore shift the $\epsilon$ required to reproduce an observed planet.

As previously noted \citep[e.g.,][]{batygin_evolution_2011}, the planetary radius depends primarily on the total injected heat rather than on the depth of energy deposition, provided the heating occurs below the radiative-convective boundary. This is illustrated in Fig.~\ref{Fig: HJ profiles}, where models with $\epsilon_{\rm inner}=2\%$ and $\epsilon_{\rm outer}=2\%$ produce nearly identical radii. Accordingly, for a given set of observed planetary parameters, the inferred heating efficiencies differ only slightly between the two prescriptions, with $\epsilon_{\rm outer}$ typically exceeding $\epsilon_{\rm inner}$ by only $0.1$--$5\%$ in relative terms, with the largest differences occurring for the most inflated planets. This small offset likely arises from differences in the location of the RCB and the heating profile: in the $\epsilon_{\rm outer}$ models, energy deposited above the 1 Mbar level can escape somewhat more efficiently, requiring slightly more heating to maintain the same radius. Consequently, the $\epsilon_{\rm outer}$ distribution closely resembles that of $\epsilon_{\rm inner}$ shown in Fig.~\ref{Fig: Heating efficiency vs flux}.

\section{Rotation correction and scaling law calibration}
\label{App: Rotation correction and scaling law calibration}

In this appendix, we assess how the integral scaling law (Eq.~\ref{Eq: B scaling law}) is affected by the rotational correction and calibrate our method against observational constraints for Jupiter. At the surface, Jupiter has a dipole field strength of $B_{\rm dip,surf}\simeq4.2$ G and a dipolarity of $f_{\rm dip}=0.75$ \citep{connerney_new_2022}, where $f_{\rm dip}=E_{\rm mag,dip}/E_{\rm mag,\ell<13}$. Downward extrapolation of the observed field to $r=0.83\,\mathrm{R_J}$ gives $B_{\rm dip}\simeq7.4$ G, $B_{\rm rms,dyn}\simeq13$ G, and $f_{\rm dip}\sim0.5$. Since magnetic energy scales as $E_{\rm mag}\propto B^2$, this dipolarity implies $B_{\rm dip}\simeq\sqrt{f_{\rm dip}}\,B_{\rm rms,dyn} \simeq B_{\rm rms,dyn}/\sqrt{2}$ at the Jovian dynamo surface. This therefore motivates adopting $K_{\rm dip}=1/\sqrt{2}$ when calibrating our prescription to Jupiter.

Previous implementations of the integral form of the \citet{christensen_energy_2009} dynamo scaling law for giant planets generally adopted the standard free-fall MLT convective velocity without explicitly accounting for rotational effects \citep[e.g.,][]{hori_linkage_2021, kilmetis_magnetic_2024, elias-lopez_planetary_2025, vigano_inflated_2025}. Applied to their respective Jupiter interior models (also using MESA), these implementations typically yield $B_{\rm rms,dyn}\simeq60$--65 G, substantially above estimates of Jupiter's dynamo field. By comparison, the simpler R09 scaling law (Eq.~\ref{Eq: Bdyn scaling law Reiners10}) predicts $B_{\rm rms,dyn}\simeq23$ G, still approximately a factor of two above the $\simeq13$ G inferred for Jupiter.

This discrepancy does not uniquely imply that the standard MLT convective velocity is overestimated, since the inferred field also depends on the adopted interior structure, composition, core structure, and dynamo-region definition. However, one possible contribution is that standard MLT does not account for the suppression of convection by rapid rotation, potentially overestimating the convective power available to the dynamo (Eq.~\ref{Eq: Convective heat flux}). We therefore introduce the analytical rotational constraint derived by \citet{fuentes_rotation_2023},
\begin{equation}
    v_{\rm conv,R}=v_{\rm conv,NR}
    \left(\frac{v_{\rm conv,NR}}{2\Omega H_P}\right)^{1/5},
\end{equation}
which relates the free-fall convective velocity predicted by standard mixing-length theory, $v_{\rm conv,NR}$, to that expected in the rapidly rotating regime. As discussed in the main text, this prescription reduces $v_{\rm conv}$ by a factor of $\sim6$ in Jupiter's interior. In our models, the correction is evaluated locally from $v_{\rm conv,NR}(r)$ and $H_P(r)$ and therefore varies throughout the convective region, typically reducing $v_{\rm conv}$ by factors of $\sim1$--5 depending primarily on $P_{\rm orb}$. The corresponding reduction in convective power lowers the predicted magnetic-field strengths by factors of $\sim3$--7, with the largest corrections occurring for the most rapidly rotating (shortest-period) planets.

Table~\ref{Tab: Jupiter calibration} summarizes the resulting estimates for Jupiter using both dynamo-radius definitions discussed in Sect.~\ref{Sec: Dynamo radius definition}. With the rotational correction, the models yield $B_{\rm rms,dyn}\simeq7$--8 G, bringing the integral scaling-law estimates to the same order as the $\simeq13$ G inferred for Jupiter, although the precise value remains sensitive to the adopted interior structure and dynamo radius. For $R_{\rm dyn}=r({\rm Rm}=50)$, together with the observationally motivated $K_{\rm dip}=1/\sqrt{2}$ derived above, the irradiated, non-inflated Jupiter model yields $B_{\rm dip,surf}\simeq4.12$ G, in close agreement with the observed $\simeq4.2$ G. We therefore adopt $K_{\rm dip}=1/\sqrt{2}$ throughout this work, while recognizing that extending this Jovian calibration to HJs remains a source of systematic uncertainty. Applying this method to the internal model of a weakly irradiated 11.9 $\mathrm{M_J}$ planet with a period of 8.5 hours at 23 Myr of age gives a $B_{\rm dip, surf}\simeq0.49$ kG (i.e., $\sim 1$ kG at the pole), which is compatible with the recent 1.25 kG measurement of $\beta$ Pictoris b by \citet{ceballos_discovery_2026}.

\begin{table}[h]
\centering
\tiny
\caption{Comparison of the magnetic field predictions for a mildly irradiated planet.}
\label{Tab: Jupiter calibration}
\begin{tabular}{cccc}
\hline\hline \\[-1.7ex]
                                    & $R_{\rm dyn}/\mathrm{R_J}$    & $v_{\rm conv, NR}$ & $v_{\rm conv, R}$ \\[.7 ex] \hline \\[-1.7ex]
\multirow{2}{*}{$B_{\rm rms,dyn}$}                                  & 0.83  & 65.35 & 7.58  \\ [.3 ex]
                                                                    & 0.93  & 58.69 & 7.21  \\ [.3 ex] \hline \\[-1.7ex]
\multirow{2}{*}{$B_{\rm dip,surf}$, $K_{\rm dip} = 1/\sqrt{2}$ }    & 0.83  & 26.56 & 3.08  \\ [.3 ex]
                                                                    & 0.93  & 34.18 & 4.12  \\ [.3 ex] \hline \hline
\end{tabular}
\tablefoot{We use a non-inflated ($\epsilon = 0$) $1 \, \mathrm{M_J}$ gas giant with $F_{\rm irr} = 7.7\times10^7$ erg cm$^{-2}$ s$^{-1}$, rotating with Jupiter's period of 9.9 h. We use both dynamo radius definitions, which for this specific model are: $R_{\rm dyn} = r(P=1\,\mathrm{Mbar}) \approx 0.83 \, \mathrm{R_J}$, $R_{\rm dyn} = r(\mathrm{Rm = 50}) \approx 0.93 \, \mathrm{R_J}$. The values shown in this table should be compared with Jupiter's $B_{\rm dip, surf}=4.2$ G, and possibly with $B_{\rm rms} (r=0.83 \, \mathrm{R_J})=13$ G.}


\end{table}

\section{Non-inflated models and threshold choice}
\label{App: Non-inflated models and threshold choice}

\begin{table}[ht]
\centering
\tiny
\caption{Age dependence of the non-inflated radius for models with $\epsilon=0$ and $F_{\rm irr}=2\times10^9\, \mathrm{erg\,cm^{-2}\,s^{-1}}$.}
\label{Tab: R0 age dependence}
\begin{tabular}{cccc}
\hline\hline \\ [-2ex]
$M_{\rm P}$ ($\mathrm{M_J}$) & $R_{\rm P}(1\,{\rm Gyr})$ ($\mathrm{R_J}$) & $R_{\rm P}(10\,{\rm Gyr})$ ($\mathrm{R_J}$) & $R_{10}/R_1$ \\
\hline \\ [-2ex]
0.5  & 1.212 & 1.075 & 0.887 \\
0.7  & 1.209 & 1.092 & 0.904 \\
1.0  & 1.206 & 1.105 & 0.916 \\
1.5  & 1.211 & 1.119 & 0.924 \\
2.0  & 1.220 & 1.131 & 0.927 \\
4.0  & 1.236 & 1.147 & 0.928 \\
8.0  & 1.229 & 1.132 & 0.921 \\
12.0 & 1.208 & 1.105 & 0.915 \\
\hline\hline
\end{tabular}
\end{table}

The reference non-inflated radius $R_{\epsilon=0}$ adopted throughout this work is evaluated at 5 Gyr. Unlike sufficiently heated models, which reach a quasi-steady state once the deposited energy largely balances the intrinsic cooling, models with $\epsilon=0$ continue to cool and contract, making $R_{\epsilon=0}$ age-dependent. We quantified this effect by evaluating the radius of representative non-inflated models between 1 and 10 Gyr. For a $1\,\mathrm{M_J}$ planet, the ratio $R_{\rm P}(10\,{\rm Gyr})/R_{\rm P}(1\,{\rm Gyr})$ remains nearly constant at $\simeq0.93$ over $10^7\leq F_{\rm irr}\leq10^9\,{\rm erg\,cm^{-2}\,s^{-1}}$, indicating a weak dependence on irradiation. The contraction depends more strongly on planetary mass: at $F_{\rm irr}=2\times10^9\,{\rm erg\,cm^{-2}\,s^{-1}}$, this ratio ranges from $\simeq0.89$ at $0.5\,\mathrm{M_J}$ to $\simeq0.93$ at intermediate masses, decreasing slightly towards the highest masses explored (see Table~\ref{Tab: R0 age dependence}). Thus, $R_{\epsilon=0}$ varies by approximately $7$--$11\%$ between 1 and 10 Gyr across these representative models.

This age dependence introduces an additional uncertainty for planets close to the non-inflated sequence, including weakly heated models for which the quasi-steady state is reached after 1 Gyr. Adopting an earlier age would increase $R_{\epsilon=0}$ and exclude more planets, whereas a later age would decrease it and slightly enlarge the interpolated sample. Given the generally poor age constraints of planetary systems, we adopt 5 Gyr as a representative population age rather than performing an age-dependent interpolation for each planet. Accordingly, $R_{\epsilon=0}$ should be regarded as a model-dependent reference radius rather than a unique boundary between inflated and non-inflated planets. This uncertainty is most relevant near the adopted radius threshold and provides additional motivation for excluding marginally inflated planets from the scaling-law calibration.

The analytical scaling laws presented in Sect.~\ref{Sec: Scaling laws} were calibrated using a restricted subset of the interpolated HJ population. As described in the main text, we excluded planets satisfying $R_{\rm P}<1.05\,R_{\epsilon=0}$. The purpose of this selection is to restrict the fits to planets for which the radius excess relative to our non-inflated models is sufficiently large that the inferred anomalous heating meaningfully affects the predicted magnetic field.

\begin{figure}[t]
\centering
\includegraphics[width=\hsize]{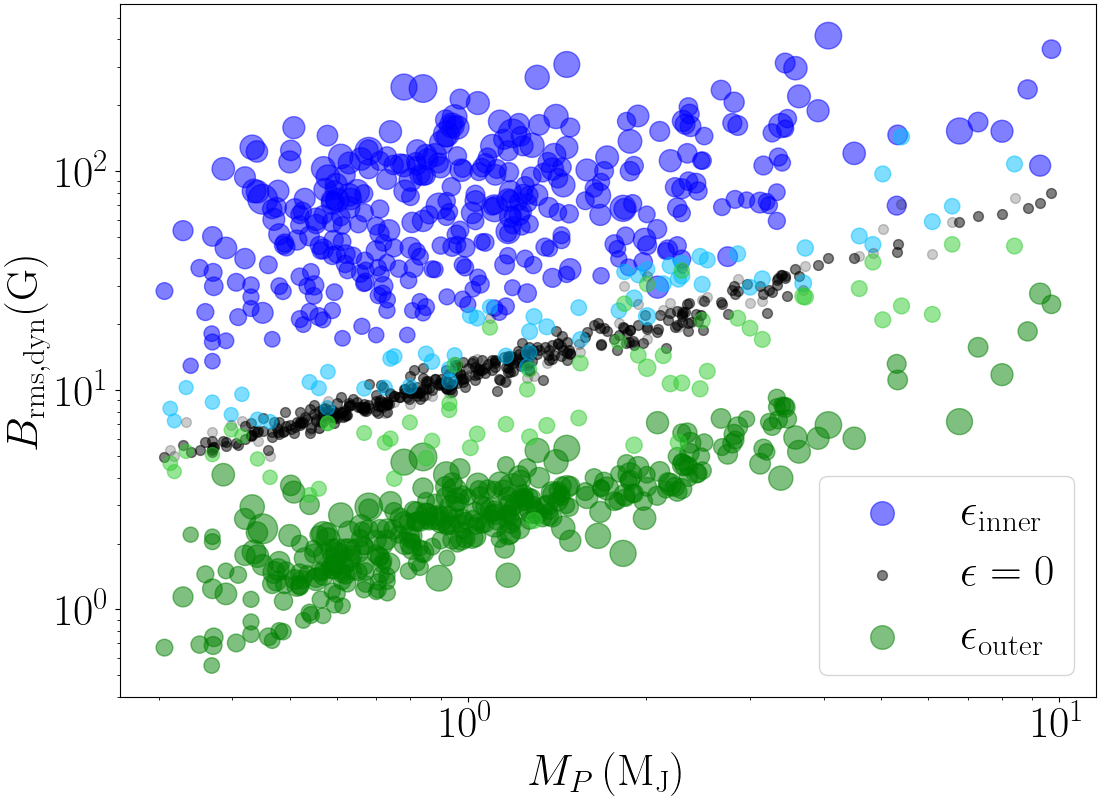}
\includegraphics[width=\hsize]{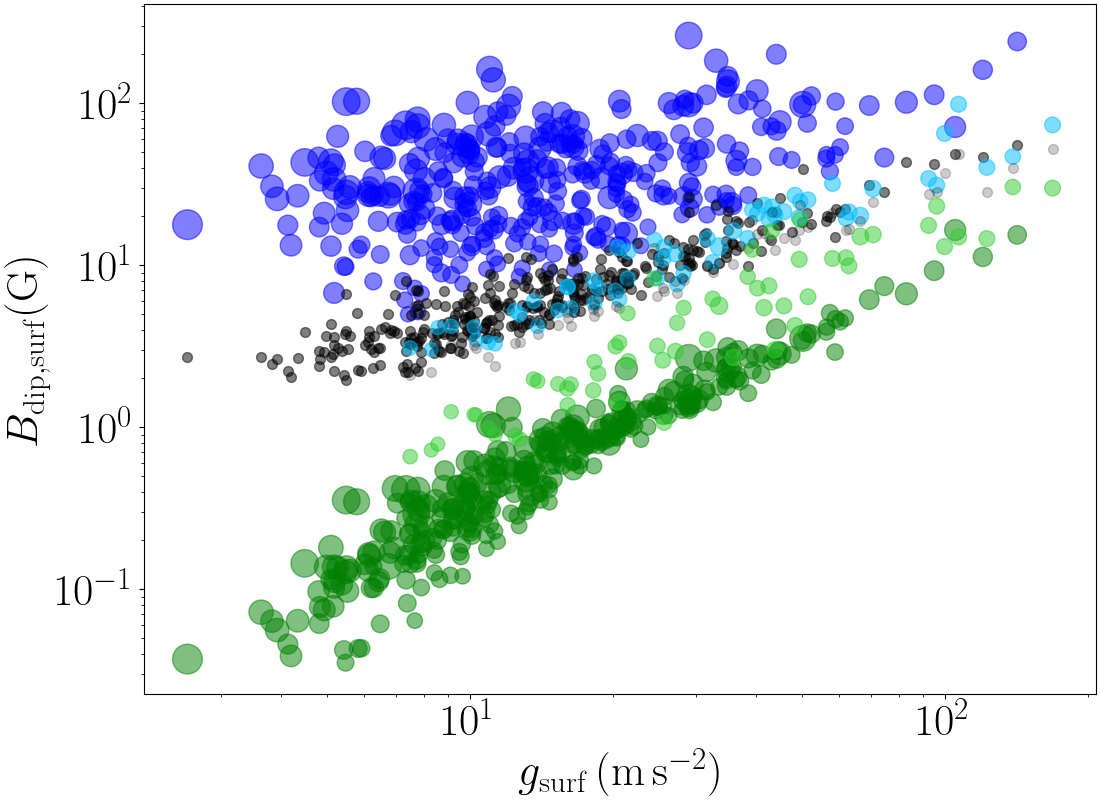}
\caption{Interpolated magnetic-field estimates for the HJ population as a function of planetary mass (top panel) and surface gravity (bottom panel). Marker size scales with the planetary radius, ranging from 1 to 2.15 $\mathrm{R_J}$. The black markers denote the interpolated values for the irradiated-only models ($\epsilon=0$) with the same $M_{\rm P}$ and $F_{\rm irr}$. In the bottom panel, all markers are placed at the surface gravity of the observed planet, $g=GM_{\rm P}/R_{\rm P}^2$. The light-colored markers indicate planets excluded from the scaling-law calibration by the criterion $R_{\rm P}<1.05\,R_{\epsilon=0}$.}
\label{Fig: Radius cut}
\end{figure} 

Figure~\ref{Fig: Radius cut} illustrates the effect of the adopted radius criterion. The lighter color symbols denote the planets excluded from the scaling-law calibration. These objects tend to populate the regions in between the non-inflated ($\epsilon=0$) sequence and the bulk of the inflated models, particularly for the $\epsilon_{\rm outer}$ prescription, where they noticeably increase the scatter. Their removal leaves the overall dependence of the magnetic field essentially unchanged, while producing significantly better power-law fits (see Fig. \ref{Fig: Fits comparisons}).

We explored several alternative selection criteria. Cuts based on the deposited heating power, $Q_{\rm dep}=\epsilon\pi R_{\rm P}^{2}F_{\rm irr}$, or on the quantity $\epsilon F_{\rm irr}$ did not isolate this transition population as effectively. A criterion based on the ratio between the magnetic fields of the two heating prescriptions, $B_{\epsilon_{\rm inner}}/B_{\epsilon_{\rm outer}}\gtrsim4$, selected nearly the same planets and produced very similar scaling relations. We nevertheless adopted the radius criterion because it is directly linked to the observable quantity used to infer the heating efficiency and does not define the sample through the magnetic field that the scaling laws are intended to predict. The value of $5\%$ should therefore be regarded as a pragmatic tolerance rather than a physical boundary between inflated and non-inflated planets.

\section{Specific exoplanet comparison}\label{App: Specific exoplanet comparison}

\begin{table*}[ht]
\centering
\tiny
\caption{Planet parameters, heating efficiency, and different $B$ estimates (all in gauss) for chosen HJs.}
\label{Tab: Specific estimates for HJs}
\begin{tabular}{cxxyyzz}
\hline \hline \\[-2.0ex]
                                                & \multicolumn{2}{c}{TOI-5322 b}  & \multicolumn{2}{c}{HATS-35 b}  & \multicolumn{2}{c}{NGTS-21 b}     \\ \hline & & & & & &  \\[-2.0ex]
 $M ({\rm M_J})$                                & \multicolumn{2}{x}{0.579$\pm$0.081}           & \multicolumn{2}{y}{1.222$\pm$0.078}           & \multicolumn{2}{z}{2.360$\pm$0.210}     \\
 $R ({\rm R_J})$                                & \multicolumn{2}{x}{1.482$\pm$0.031}           & \multicolumn{2}{y}{1.464$\pm$0.069}           & \multicolumn{2}{z}{1.330$\pm$0.030}     \\
 $F_{\rm irr}$($10^{9}$erg\,cm$^{-2}$\,s$^{-1}$)& \multicolumn{2}{x}{0.754$\pm$0.060}           & \multicolumn{2}{y}{3.88$\pm$0.40}             & \multicolumn{2}{z}{0.768$\pm$0.083}     \\ [.7 ex]
 $P_{\rm orb}$(days)                            & \multicolumn{2}{x}{5.423321$\pm$4$\times10^{-6}$}     & \multicolumn{2}{y}{1.820999$\pm$0.000002}     & \multicolumn{2}{z}{1.543390$\pm$0.000002}     \\ [.7 ex]
 $\epsilon_{\mathrm{inner}}(\%)$                & \multicolumn{2}{x}{1.37$_{-0.60}^{+1.2}$}    & \multicolumn{2}{y}{0.058$_{-0.025}^{+0.045}$} & \multicolumn{2}{z}{4.6$_{-1.9}^{+3.3}$} \\ [.7 ex]
 $\epsilon_{\mathrm{outer}}(\%)$                & \multicolumn{2}{x}{1.39$_{-0.62}^{+1.1}$}    & \multicolumn{2}{y}{0.060$_{-0.026}^{+0.047}$} & \multicolumn{2}{z}{4.8$_{-1.9}^{+3.5}$} \\ [.7 ex]
 $\epsilon_{\mathrm{Th18}}(\%)$                 & \multicolumn{2}{x}{1.84$_{-0.42}^{+0.90}$}    & \multicolumn{2}{y}{1.14$_{-0.41}^{+0.59}$}    & \multicolumn{2}{z}{1.87$_{-0.40}^{+0.91}$} \\ [.7 ex]
\hline  & & & & & & \\[-2.0ex]
 $\epsilon = \epsilon_{\rm inner}$       & $B_{\rm rms, dyn}$ & $B_{\rm dip, surf}$ &  $B_{\rm rms, dyn}$ & $B_{\rm dip, surf}$ & $B_{\rm rms, dyn}$ & $B_{\rm dip, surf}$ \\ \hline & & & & & & \\[-1.3ex]
 Interpolator                                   & 145$_{-31}^{+34}$       & 63.0$_{-16}^{+19}$        & 63.2$_{-11}^{+14}$         & 32.3$_{-5.6}^{+6.9}$       & 197$_{-35}^{+38}$           & 124$_{-22}^{+24}$ \\ [.7 ex]
 MRL fit                                        & 94.2$_{-31}^{+44}$      & 40.1$_{-13}^{+19}$        & 73.2$_{-22}^{+31}$         & 36.9$_{-11}^{+15}$       & 139$_{-45}^{+66}$           & 85.8$_{-28}^{+42}$ \\ [.7 ex]
 MRF fit                                        & 68.4$_{-16}^{+19}$      & 29.6$_{-6.8}^{+8.8}$        & 75.5$_{-18}^{+24}$         & 37.4$_{-9.0}^{+12}$       & 99.6$_{-19}^{+25}$          & 62.3$_{-12}^{+16}$ \\ [.7 ex]
 MR fit                                         & 62.4$_{-12}^{+15}$      & 26.5$_{-5.2}^{+6.6}$      & 82.7$_{-16}^{+20}$         & 41.8$_{-8.2}^{+10}$       & 88.5$_{-16}^{+18}$          & 53.9$_{-9.4}^{+11}$ \\ [.7 ex]
 R09($\epsilon_{\mathrm{inner}}$)               & 121$_{-69}^{+86}$       & 30$_{-18}^{+22}$          & 83$_{-47}^{+57}$         & 37$_{-21}^{+26}$         & 244$_{-140}^{+170}$         & 138$_{-78}^{+96}$ \\ [.7 ex]
 R09($\epsilon_{\mathrm{Th18}}$)                & 133$_{-74}^{+90}$       & 33$_{-19}^{+23}$          & 223$_{-130}^{+150}$        & 101$_{-57}^{+67}$        & 180$_{-100}^{+120}$         & 102$_{-57}^{+68}$ \\ [.7 ex]
\hline & & & & & & \\[-2.0ex]
 $\epsilon = \epsilon_{\rm outer}$       & $B_{\rm rms, dyn}$ & $B_{\rm dip, surf}$ &  $B_{\rm rms, dyn}$ & $B_{\rm dip, surf}$ & $B_{\rm rms, dyn}$ & $B_{\rm dip, surf}$ \\ \hline & & & & & & \\[-1.3ex]
 Interpolator                                   & 2.23$_{-0.49}^{+0.76}$      & 0.228$_{-0.11}^{+0.13}$  & 2.34$_{-0.13}^{+0.25}$    & 0.589$_{-0.073}^{+0.083}$  & 4.51$_{-0.28}^{+0.29}$      & 2.02$_{-0.19}^{+0.18}$ \\ [.7 ex]
  MRL fit                                        & 2.05$_{-0.66}^{+1.0}$     & 0.226$_{-0.096}^{+0.17}$ & 2.93$_{-0.86}^{+1.2}$      & 0.656$_{-0.24}^{+0.39}$  & 5.48$_{-1.8}^{+2.9}$        & 2.31$_{-0.95}^{+1.7}$ \\ [.7 ex]
 MRF fit                                        & 1.82$_{-0.43}^{+0.56}$     & 0.221$_{-0.074}^{+0.11}$& 3.08$_{-0.73}^{+0.92}$    & 0.673$_{-0.21}^{+0.32}$  & 4.76$_{-0.96}^{+1.2}$       & 2.13$_{-0.61}^{+0.86}$ \\ [.7 ex]
 MR fit                                         & 1.82$_{-0.39}^{+0.46}$     & 0.217$_{-0.068}^{+0.092}$& 3.09$_{-0.59}^{+0.71}$    & 0.681$_{-0.18}^{+0.23}$  & 4.74$_{-0.89}^{+1.1}$      & 2.07$_{-0.54}^{+0.71}$ \\ [.7 ex]
\hline & & & & & & \\[-2.0ex]
 $\epsilon = 0$                           & $B_{\rm rms, dyn}$ & $B_{\rm dip, surf}$ &  $B_{\rm rms, dyn}$ & $B_{\rm dip, surf}$ & $B_{\rm rms, dyn}$ & $B_{\rm dip, surf}$ \\ \hline & & & & & & \\[-1.3ex]
 Interpolator                                   & 7.96$_{-0.86}^{+0.87}$      & 3.96$_{-0.61}^{+0.60}$    & 13.4$_{-0.7}^{+0.6}$      & 7.3$_{-0.5}^{+0.5}$     & 18.7$_{-1.2}^{+1.2}$        & 11.9$_{-0.9}^{+0.9}$   \\ [.7 ex]
 MR fit                                         & 8.18$_{-1.2}^{+1.3}$      & 3.93$_{-0.63}^{+0.70}$    & 14.0$_{-1.6}^{+1.8}$      & 7.6$_{-0.9}^{+1.1}$     & 22.3$_{-2.6}^{+2.9}$        & 13.8$_{-1.7}^{+2.0}$ \\ [.7 ex]
\hline \hline
\end{tabular}
\tablefoot{The $\epsilon_{\rm inner}$ and $\epsilon_{\rm outer}$ heating efficiencies are determined directly from interpolation. The row named ``MRL fit'' refers to Eq.~\ref{Eq: Scaling law MRL}, ``MRF fit'' to Eq.~\ref{Eq: Scaling law MRF}, ``MR fit'' to Eq.~\ref{Eq: Scaling law MR},  R09($\epsilon_{\mathrm{inner}}$) refers to Eq.~\ref{Eq: Bdyn scaling law Reiners10} \citep{reiners_surprisingly_2009} using $L_{\rm P} = \epsilon_{\rm inner} \pi R_P^2F_{\rm irr}$, and R09($\epsilon_{\mathrm{Th18}}$) refers to the same expression but using $L_{\rm P} = \epsilon_{\rm Th18} \pi R_P^2F_{\rm irr}$.}
\end{table*}

To illustrate the performance of the proposed scaling laws relative to previous prescriptions, Table~\ref{Tab: Specific estimates for HJs} compares magnetic field estimates for three representative inflated HJs ($R_{\rm P}>1.3\,\mathrm{R_J}$). The planets were selected such that their interpolated $\epsilon$ values lie below, near, and above the corresponding $\epsilon_{\rm Th18}$ values (Eq.~34 of \citealt{thorngren_bayesian_2018}). For each planet and heating prescription, we compare the interpolated values with the MRL, MRF, and MR relations derived above and with the R09 scaling law.

For $\epsilon_{\rm inner}$, our scaling relations generally reproduce the interpolated magnetic fields within their uncertainties. Some exceptions are the MRF and MR scaling-law estimates for TOI-5322 b and NGTS-21 b, which fall outside $1\sigma$ of the interpolator. The MRL relation provides the closest overall agreement (as seen by the Pearson $r$ closest to unity), although the simpler MRF and MR relations remain satisfactory. The R09 scaling law with our interpolated $\epsilon_{\rm inner}$ also gives estimates consistent with the fits and interpolator, and so does R09 with $\epsilon_{\rm Th18}$ for TOI-5322 b and NGTS-21 b. For HATS-35 b, whose interpolated $\epsilon$ is $\sim$20 times lower than $\epsilon_{\rm Th18}$, R09($\epsilon_{\rm Th18}$) overestimates the field by a factor of $\sim$3. Similarly, for $\epsilon_{\rm outer}$, the MRL, MRF, and MR relations remain in good agreement with the interpolator despite the substantially weaker predicted fields. These examples show that the proposed scaling laws provide practical approximations to the full interpolation while requiring only a small number of parameters.

\section{Comparison with literature values}\label{App: Comparison with literature values}

To compare our magnetic-field predictions with the indirect constraints, we adopt the planetary parameters reported in their respective studies. The values of $M_{\rm P}$, $R_{\rm P}$, and $F_{\rm irr}$ (the latter calculated from $R_\star$, $T_{\rm eff}$, and $a$) are listed in Table~\ref{Tab: Literature comparison planet parameters}. Although some differ slightly from current values in the NASA Exoplanet Archive, we use the published parameters for consistency with the original analyses. Table~\ref{Tab: Literature comparison} contains the magnetic field estimates for these planets. Figure~\ref{Fig: B comparison with literature} shows the interpolated estimates, except for $\tau$~Boo~b, for which the MR scaling law is used for the reasons discussed above.

Three systems analyzed by \citet{cauley_magnetic_2019} (HD~179949~b, $\tau$~Boo~b, and $\upsilon$~And~b) are non-transiting and therefore lack measured radii. We adopt the radii inferred by \citet{cauley_magnetic_2019} from the mass--radius relation of \citet{chen_probabilistic_2016}. HD~179949~b and $\upsilon$~And~b fall within our interpolation grid, whereas the inferred radius of the more massive $\tau$~Boo~b lies below its lower boundary; for this planet we therefore use the MR scaling law. For the SPI-derived magnetic fields, we show the values corresponding to their nominal, but poorly constrained, efficiency $E_{\rm CaIIK}/E_{\rm tot}=0.20\%$.

For the planets analyzed by \cite{seidel_magnetic_2026}, we likewise adopt the stellar and planetary parameters listed in their tables, as these are the best estimates coming from spectroscopic measurements, which are used to infer the atmospheric wind speeds and magnetic field strengths. Beware that such inferred values are upper limits, since the local magnetic field includes both the dynamo-generated field and the induced one, with the latter potentially dominating.

\begin{table*}[ht]
\centering
\tiny
\caption{Planetary parameters used in Table~\ref{Tab: Literature comparison}.}
\label{Tab: Literature comparison planet parameters}
\begin{tabular}{ccccc}
\hline \hline \\[-2.0ex]
 Planet & $M_{\rm P}(\mathrm{M_J})$ & $R_{\rm P}(\mathrm{R_J})$ & $F_{\rm irr}$($10^{9}$erg\,cm$^{-2}$\,s$^{-1}$) & $P_{\rm orb}$(days) \\ \hline  \\[-2.0ex]
WASP-12b      & 1.41$\pm$0.10 & 1.79$\pm$0.09 & 9.08$\pm$3.5      & 1.091423$\pm$3.0$\times10^{-6}$ \\
HD209458b     & 0.714$\pm$0.017 & 1.380$\pm$0.017 & 1.029$\pm$0.048 & 3.524749$\pm$3.8$\times10^{-7}$ \\
HD189733b     & 1.14$\pm$0.03 & 1.14$\pm$0.03 & 0.467$\pm$0.025     & 2.2186$\pm$1.5$\times10^{-7}$ \\
HD179949b*    & 1.04$\pm$0.08 & 1.22$\pm$0.18 & 1.358$\pm$0.085     & 3.0925$\pm$3.2$\times10^{-5}$ \\
$\tau$ Boo b* & 6.13$\pm$0.34 & 1.13$\pm$0.17 & 1.88$\pm$0.21       & 3.3124$\pm$6.9$\times10^{-6}$ \\
$\upsilon$ And b*  & 0.69$\pm$0.06 & 1.25$\pm$0.19 & 1.407$\pm$0.087     & 4.6170$\pm$2.3$\times10^{-5}$ \\
WASP-76b      & 0.92$\pm$0.03 & 1.83$\pm$0.06 & 5.571$\pm$0.51      & 1.8099$\pm$6.4$\times10^{-7}$ \\
KELT-20b      & 2.16$\pm$0.21 & 1.83$\pm$0.07 & 5.912$\pm$0.97      & 3.4741$\pm$1.9$\times10^{-6}$ \\
WASP-121b     & 1.18$\pm$0.07 & 1.87$\pm$0.04 & 7.642$\pm$0.31      & 1.2749$\pm$1.5$\times10^{-7}$ \\
WASP-178b     & 1.41$\pm$0.40 & 1.94$\pm$0.06 & 8.431$\pm$0.94      & 3.3448$\pm$1.2$\times10^{-6}$ \\
TOI-1518b     & 1.83$\pm$0.47 & 1.88$\pm$0.04 & 9.527$\pm$0.79      & 1.9026$\pm$4.3$\times10^{-7}$ \\
HAT-P-70b     & 1.66$\pm$0.22 & 1.87$\pm$0.15 & 9.611$\pm$3.40      & 2.7443$\pm$7.9$\times10^{-7}$ \\
WASP-189b     & 1.99$\pm$0.16 & 1.62$\pm$0.02 & 10.96$\pm$0.67      & 2.7240$\pm$2.0$\times10^{-6}$ \\
\hline \hline 
\end{tabular}
\tablefoot{The planets marked with * do not have radius measurements and the reported values of $R_{\rm P}$ are derived using the mass-radius relation given by \citet{chen_probabilistic_2016}.}
\end{table*}

\begin{table*}[ht]
\centering
\tiny
\caption{Comparison between published observational constraints on HJ magnetic fields and our predictions.}
\label{Tab: Literature comparison}
\begin{tabular}{ccbbvvgg}
\hline \hline & & & & & & & \\[-2.0ex]
\multirow{2}{*}{Planet} &
\multirow{2}{*}{Literature value} &
\multicolumn{2}{b}{$B_{\rm dip,surf},\,\epsilon_{\rm inner}$} & \multicolumn{2}{v}{$B_{\rm dip,surf},\,\epsilon_{\rm outer}$} & \multicolumn{2}{g}{$B_{\rm dip,surf},\,\epsilon=0$}\\[0.5ex]
& & Interp.&MR fit& Interp.&MR fit& Interp.&MR fit\\ 

\hline  \\[-1.5ex]
\multicolumn{8}{c}{\citep{vidotto_early_2010}: maximum polar magnetic field, $B_{\rm pole, surf, max}$}\\[.5 ex]
\hline & & & & & & & \\[-1.5ex]
WASP-12 b & 24 & 23.6$_{-9.1}^{+17}$ & 65$_{-15}^{+21}$ & 0.33$_{-0.08}^{+0.19}$ & 0.66$_{-0.22}^{+0.31}$ & 7.47$_{-0.62}^{+0.62}$ & 8.6$_{-1.3}^{+1.6}$\\[.5 ex]

\hline \\[-1.5ex]
\multicolumn{8}{c}{\citep{kislyakova_magnetic_2014}: surface dipolar magnetic field, $B_{\rm dip, surf}$}\\[.5 ex]
\hline & & & & & & & \\[-1.5ex]
HD 209458 b & 0.16 & 52.6$_{-4.2}^{+4.5}$ & 26.8$_{-4.5}^{+5.3}$ & 0.361$_{-0.026}^{+0.027}$ & 0.326$_{-0.079}^{+0.096}$ & 4.56$_{-0.12}^{+0.12}$ & 4.75$_{-0.50}^{+0.55}$\\[.5 ex]

\hline \\[-1.5ex]
\multicolumn{8}{c}{\citep{cauley_magnetic_2019}: polar magnetic field, $B_{\rm pole, surf}$}\\[.5 ex]
\hline & & & & & & & \\[-1.5ex]
HD189733b & 20$\pm$7 & 14.9$_{-5.1}^{+6.6}$ & 26.0$_{-3.5}^{+4.1}$ & 0.812$_{-0.059}^{+1.5}$ & 0.837$_{-0.16}^{+0.19}$ & 6.64$_{-0.16}^{+0.16}$ & 7.29$_{-0.58}^{+0.62}$\\[.5 ex]
HD179949b* & 86$\pm$29 & 25$_{-18}^{+48}$ & 27.5$_{-6.9}^{+9.3}$ & 0.85$_{-0.16}^{+3.9}$ & 0.67$_{-0.19}^{+0.24}$ & 6.63$_{-0.53}^{+0.50}$ & 6.7$_{-0.8}^{+0.8}$\\[.5 ex]
$\tau$ Boo b* & 117$\pm$38 & -- & 75$_{-21}^{+28}$ & -- & 10.7$_{-3.4}^{+4.6}$ & -- & 32.5$_{-4.4}^{+4.9}$\\[.5 ex]
$\upsilon$ And b* & 83$\pm$77 & 21$_{-16}^{+30}$ & 22.1$_{-6.0}^{+7.7}$ & 0.38$_{-0.11}^{+2.2}$ & 0.35$_{-0.11}^{+0.13}$ & 4.60$_{-0.46}^{+0.43}$ & 4.64$_{-0.55}^{+0.60}$\\[.5 ex]

\hline \\[-1.5ex]
\multicolumn{8}{c}{\citep{turner_search_2021}: maximum polar magnetic field, $B_{\rm pole, surf, max}$}\\[.5 ex]
\hline & & & & & & & \\[-1.5ex]
$\tau$ Boo b* & 8$\pm$3 & -- & 75$_{-21}^{+28}$ & -- & 10.7$_{-3.4}^{+4.6}$ & -- & 32.5$_{-4.4}^{+4.9}$\\[.5 ex]

\hline \\[-1.5ex]
\multicolumn{8}{c}{\citep{seidel_magnetic_2026}: maximum atmospheric magnetic field, $B_{\rm atm, max}$}\\[.5 ex]
\hline & & & & & & & \\[-1.5ex]
WASP-76 b & 1.74$_{-0.11}^{+0.12}$ & 28.0$_{-3.0}^{+3.0}$ & 51$_{-12}^{+15}$ & 0.151$_{-0.020}^{+0.022}$ & 0.336$_{-0.11}^{+0.16}$ & 5.2$_{-0.2}^{+0.2}$ & 5.8$_{-0.8}^{+1.0}$\\[.5 ex]
KELT-20b & 2.14$_{-0.46}^{+0.69}$ & 74$_{-13}^{+16}$ & 89$_{-22}^{+29}$ & 0.84$_{-0.14}^{+0.15}$ & 1.22$_{-0.42}^{+0.63}$ & 14.3$_{-1.3}^{+1.2}$ & 12.5$_{-2.0}^{+2.5}$\\[.5 ex]
WASP-121 b & 2.15$_{-0.19}^{+0.23}$ & 25.8$_{-2.3}^{+2.3}$ & 63$_{-14}^{+19}$ & 0.186$_{-0.032}^{+0.033}$ & 0.48$_{-0.16}^{+0.24}$ & 6.4$_{-0.4}^{+0.4}$ & 7.3$_{-1.1}^{+1.3}$\\[.5 ex]
WASP-178 b & 2.00$_{-0.43}^{+0.67}$ & 35.2$_{-13}^{+12}$ & 75$_{-21}^{+29}$ & 0.265$_{-0.13}^{+0.17}$ & 0.60$_{-0.28}^{+0.44}$ & 9.5$_{-3.0}^{+2.5}$ & 8.5$_{-2.4}^{+2.8}$\\[.5 ex]
TOI-1518b & 2.18$_{-0.27}^{+0.35}$ & 34.0$_{-10}^{+8.1}$ & 83$_{-23}^{+32}$ & 0.40$_{-0.15}^{+0.22}$ & 0.92$_{-0.39}^{+0.64}$ & 10.8$_{-2.7}^{+2.5}$ & 10.8$_{-2.9}^{+3.2}$\\[.5 ex]
HAT-P-70 b & 3.5$_{-1.2}^{+3.3}$ & 32.0$_{-12}^{+23}$ & 78$_{-20}^{+30}$ & 0.47$_{-0.16}^{+0.27}$ & 0.80$_{-0.30}^{+0.44}$ & 10.5$_{-1.4}^{+1.3}$ & 9.9$_{-1.8}^{+2.1}$\\[.5 ex]
WASP-189 b & 5.5$_{-2.7}^{+13.0}$ & 26.2$_{-3.1}^{+3.0}$ & 68$_{-14}^{+18}$ & 0.787$_{-0.096}^{+0.12}$ & 1.25$_{-0.39}^{+0.53}$ & 12.5$_{-0.9}^{+0.9}$ & 11.7$_{-1.7}^{+1.9}$\\ [.5 ex]

\hline \hline 
\end{tabular}
\tablefoot{The uncertainties listed in this table only include the propagation of the observational errors (planetary and stellar parameters), and they do not include the systematic uncertainty associated with the planetary evolution models and the dynamo scaling law. For the comparison shown in Fig.~\ref{Fig: B comparison with literature}, an additional factor of 2 relative systematic uncertainty is added in quadrature.}
\end{table*}

\section{Interior model caveats}
\label{App: Interior model caveats}

Our magnetic-field estimates inherit the assumptions of the underlying planetary evolution models. In particular, we do not include hydrogen-helium phase separation (helium rain). Interior models of Jupiter and Saturn suggest that helium rain can produce stably stratified layers (SSLs) near the $\sim$Mbar pressure regime \citep{stevenson_phase_1977, markham_stable_2024, tejada_arevalo_jupiter_2025}, potentially modifying both the transport of internal heat and the dynamo behavior \citep[e.g.,][]{gastine_stable_2021, moore_dynamo_2022}. This effect is expected to be less important for the highly irradiated HJs that are the primary focus of this work, although it could become relevant for the cooler or more weakly inflated planets in our sample. Likewise, recent studies favor diluted or \textit{fuzzy} cores rather than a sharp core-envelope boundary, implying composition gradients extending over a significant fraction of the planetary interior \citep[e.g.,][]{wahl_comparing_2017, mankovich_diffuse_2021, militzer_juno_2022, helled_fuzzy_2024, sur_simultaneous_2025}. In contrast, our MESA models assume adiabatic, fully convective, chemically homogeneous envelopes surrounding a distinct inert core. The possible impact of diluted cores and composition gradients on the magnetic fields of HJs, together with the role of helium rain in the cooler members of the population, remains an important topic for future investigation.

As mentioned in Sect.~\ref{Sec: Simulation grid and interpolator}, the interpolator was constructed using the evolved models at 5 Gyr. Previous studies of HJ inflation have shown that the intrinsic convective luminosity associated with secular cooling is much smaller than the power required to sustain radius inflation \citep[e.g.][]{thorngren_bayesian_2018,sarkis_evidence_2021}. Our choice of a single representative age therefore relies on the assumption that the heating efficiency remains approximately constant over time and is sufficiently large to balance the secular cooling, producing the radius plateau observed in our models after $\sim$0.1--1 Gyr. Under these assumptions, the internal heat deposition dominates the energy budget and the planetary radius becomes only weakly dependent on age. To show how this approximation affects our results, Table~\ref{Tab: Age dependence} compares magnetic field estimates for the same models at different evolutionary stages. For both the $\epsilon_{\rm inner}$ and $\epsilon_{\rm outer}$ prescriptions, the predicted values of $B_{\rm rms,dyn}$ and $B_{\rm dip,surf}$ remain nearly constant between 1 and 10 Gyr. This indicates that the deposited heat controls the convective heat flux entering Eq.~\ref{Eq: B scaling law}. Consequently, the choice of 5 Gyr has a negligible impact on the inferred magnetic field strengths of inflated HJs within the framework of our models. In contrast, the non-inflated models ($\epsilon=0$) exhibit the expected decline in magnetic field strength with age as the planet cools and its intrinsic luminosity decreases \citep[e.g.\ $B\propto t^{-0.2\text{-}0.3}$,][]{reiners_magnetic_2010,elias-lopez_planetary_2025}. This should be taken into account when assessing young giant planets ($\lesssim$0.1 Gyr), as the primordial inflated radii and strong formation heat can drive vigorous convection. Our estimates and scaling laws are therefore not expected to apply to such young systems. Moreover, our calculations assume a constant incident stellar irradiation throughout the evolution as well as the final stages of the stellar evolution. In reality, stellar luminosities evolve with time, so the irradiation history may modify the long-term thermal planetary evolution. Nevertheless, since most well-characterized HJs orbit mature main-sequence stars and are several Gyr old, the assumptions of constant heating efficiency and irradiation should provide a practical approximation for the majority of the currently observed population.

\begin{table}[t]
\centering
\tiny
\caption{Magnetic field estimates at different evolutionary times (1 $\mathrm{M_J}$, $F_{\rm irr} = 10^8 \, \mathrm{erg \, cm^{-2} \, s^{-1}}$, $P_{\rm orb}=2$ days).}
\label{Tab: Age dependence}
\begin{tabular}{cbvg}
\hline \hline \\[-2.0ex]
      \multicolumn{4}{c}{$B_{\rm rms, dyn}$ (G)}      \\ \hline & & & \\ [-2.0ex]
 Age (Gyr)       & $\epsilon_{\rm inner}$ & $\epsilon_{\rm outer}$ &  $\epsilon = 0$ \\ \hline & & &\\[-2.0ex]
 1 Gyr           & 63.41       & 2.37        & 20.08        \\ 
 2 Gyr           & 63.41       & 2.28        & 14.35        \\ 
 5 Gyr           & 63.41       & 2.13        & 9.97        \\ 
 10 Gyr          & 63.41       & 1.97        & 7.41       \\ 
\hline \\[-2.0ex]
     \multicolumn{4}{c}{$B_{\rm dip, surf}$ (G)}      \\ \hline  & & & \\ [-2.0ex]
 Age (Gyr)       & $\epsilon_{\rm inner}$ & $\epsilon_{\rm outer}$ &  $\epsilon = 0$ \\ \hline & & & \\[-2.0ex]
 1 Gyr           & 35.45       & 0.37        & 11.45        \\ 
 2 Gyr           & 35.45       & 0.36        & 8.19       \\ 
 5 Gyr           & 35.45       & 0.33        & 5.76        \\ 
 10 Gyr          & 35.45       & 0.31        & 4.32        \\
\hline \hline 
\end{tabular}
\end{table}

Another possible source of uncertainty is the assumed EoS used to close the structure equations (Eqs.~\ref{Eq: MESA mass conservation} - \ref{Eq: MESA energy transport}). Throughout this work, we have adopted MESA's standard SCvH H-He EoS \citep{saumon_equation_1995}, although alternative formulations based on more recent \textit{ab initio} calculations are available \citep[e.g.][]{militzer_ab_2013, chabrier_new_2019, chabrier_new_2021, mazevet_benchmarking_2022, howard_accounting_2023}. Different EoS prescriptions give slightly different $\rho-T$ relations and therefore modify internal properties such as planetary radius and total metallicity \citep{miguel_jupiter_2016}, which can affect the convective flux and thus our magnetic field estimates (Eq. \ref{Eq: Convective heat flux}). To assess the magnitude of this effect, we follow \citet{helled_giant_2025} and evolve HJs using the more recent CD21 EoS \citep{chabrier_new_2021}, while varying assumptions about the heavy-element component. The results are shown in Table~\ref{Tab: EOS Brms estimates}. The largest difference is found between a pure H-He EoS and EoS tables including heavy elements, which reduce the predicted magnetic field strengths by $\sim20$--$30\%$ for the inflated models. In contrast, once heavy elements are included, the choice between the SCvH and CD21 formulations produces only minor changes, typically of order a few percent. Similarly, assuming a pure H$_2$O heavy-element component or a 50\% H$_2$O / 50\% rock mixture has a negligible impact on the resulting field strengths. These results indicate that our magnetic field estimates are relatively insensitive to the specific EoS prescription adopted, provided that the heavy-element contribution is treated consistently. 

\begin{table}[ht]
\centering
\tiny
\caption{Magnetic field estimates for different equations of state (1 $\mathrm{M_J}$, $F_{\rm irr} = 5 \times 10^8 \, \mathrm{erg \, cm^{-2} \, s^{-1}}$, $Z$=0.02, $P_{\rm orb}=2$ days).}
\label{Tab: EOS Brms estimates}
\begin{tabular}{cbvg}
\hline \hline \\[-2.0ex]
      \multicolumn{4}{c}{$B_{\rm rms, dyn}$ (G)}      \\ \hline & & & \\ [-2.0ex]
 EoS       & $\epsilon_{\rm inner}$ & $\epsilon_{\rm outer}$ &  $\epsilon = 0$ \\ \hline & & &\\[-2.0ex]
 SCvH                       & 125.9       & 3.029       & 10.06        \\ 
 SCvH + H$_2$O              & 101.5       & 2.213       & 9.305        \\ 
 SCvH, 50\%H$_2$O/50\% Rock & 101.6       & 2.217       & 9.318       \\ 
 CD21, H$_2$O               & 100.3       & 1.892       & 9.333      \\ 
 CD21, 50\%H$_2$O/50\% Rock & 100.4       & 1.896       & 9.372      \\ 
\hline \\[-2.0ex]
     \multicolumn{4}{c}{$B_{\rm dip, surf}$ (G)}      \\ \hline  & & & \\ [-2.0ex]
 EoS       & $\epsilon_{\rm inner}$ & $\epsilon_{\rm outer}$ &  $\epsilon = 0$ \\ \hline & & & \\[-2.0ex]
 SCvH                       & 34.58       & 0.3990      & 2.886        \\ 
 SCvH + H$_2$O              & 27.59       & 0.2886      & 2.647        \\ 
 SCvH, 50\%H$_2$O/50\% Rock & 27.63       & 0.2896      & 2.656       \\ 
 CD21, H$_2$O               & 27.70       & 0.2737      & 2.707      \\ 
 CD21, 50\%H$_2$O/50\% Rock & 27.75       & 0.2747      & 2.722      \\ 
\hline \hline 
\end{tabular}
\end{table}

The total heavy-element abundance, $Z_{\rm tot}$, represents another important source of uncertainty. Mass--metallicity relations ($M_{\rm P}$--$M_Z$) inferred from well-characterized, non-inflated giant planets indicate that bulk metallicity decreases with planetary mass on a population level, such that lower-mass planets are generally more metal-enriched \citep[e.g.,][]{thorngren_mass-metallicity_2016, muller_mass-radius_2024}. HJs are expected to follow a similar trend, but the large intrinsic scatter prevents these relations from providing reliable metallicity estimates for individual planets. Moreover, metallicity is strongly degenerate with heating efficiency: the same observed radius can often be reproduced by different combinations of $Z$ and $\epsilon$ \citep{thorngren_bayesian_2018, sarkis_evidence_2021}. This degeneracy introduces an additional uncertainty in the inferred interior structure and, consequently, in the predicted magnetic field. In App.~\ref{App: Z dependence}, we quantify how uncertainties in the assumed HJ metallicity propagate into our magnetic-field estimates.

Taken together, these model assumptions and order-unity uncertainties are difficult to quantify and propagate individually and are not necessarily independent. We therefore adopt a representative relative systematic uncertainty within a factor of 2 on our magnetic-field estimates, comparable to that assumed by \citet{reiners_surprisingly_2009} for their magnetic-field scaling relation. This value is intended to encompass the combined uncertainties associated with the interior structure and dynamo scaling law, and should therefore be interpreted as an approximate systematic uncertainty rather than a formal statistical confidence interval.

\subsection{Metallicity dependence}\label{App: Z dependence}

\begin{figure}[t]
\centering
\includegraphics[width=.9\hsize]{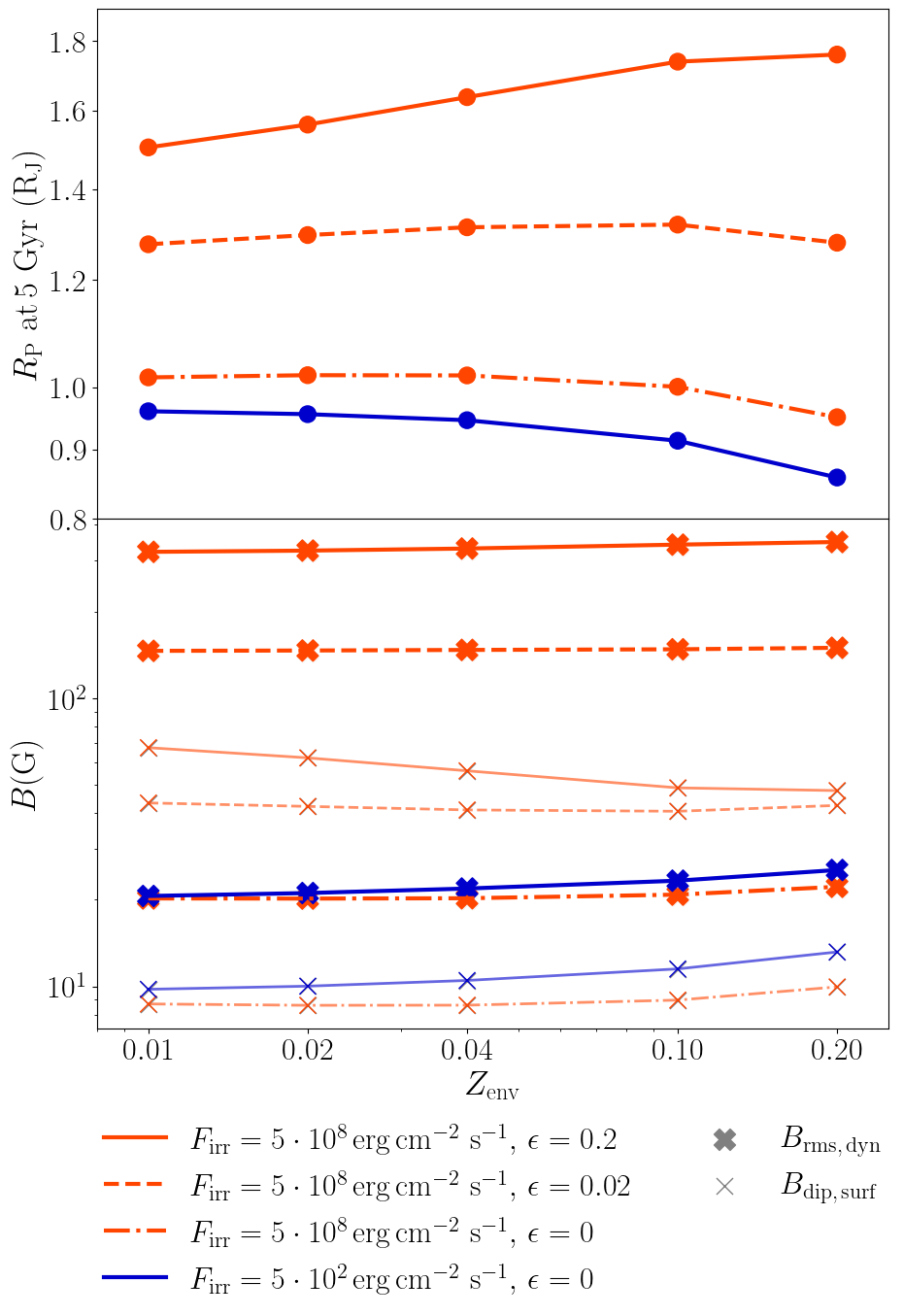}
\caption{Planetary radius (top) and magnetic field strengths (bottom) at 5 Gyr as a function of envelope metallicity, $Z_{\rm env}$, for $1\,\mathrm{M_J}$ models evolved with the $\epsilon_{\rm inner}$ heating prescription.}
\label{Fig: R and Brms vs Z}
\end{figure}

To investigate how uncertainties in the envelope metallicity ($Z$) affect our magnetic-field predictions, we computed additional $1\,\mathrm{M_J}$ evolutionary models spanning $0.01\leq Z\leq0.20$, while keeping the H/He ratio fixed and maintaining the $10\,M_{\oplus}$ inert core. The helium abundance was adjusted as $Y=(1-Z)/(K+1)$, with $K=X/Y$ constant. All models were evolved with the $\epsilon_{\rm inner}$ prescription, and their radii and magnetic fields were evaluated at 5 Gyr. Figure~\ref{Fig: R and Brms vs Z} shows $R_{\rm P}$, $B_{\rm rms,dyn}$, and $B_{\rm dip,surf}$ for representative irradiation levels and heating efficiencies.

For non-inflated planets ($\epsilon_{\rm inner}=0$), increasing $Z$ generally produces smaller radii because of the larger mean molecular weight, consistent with previous giant-planet evolution models \citep[e.g.,][]{thorngren_mass-metallicity_2016}. This trend weakens with inflation and reverses for the most inflated models ($R_{\rm P}\gtrsim1.5\,\mathrm{R_J}$), where metal-rich envelopes become larger. This reflects competition between the radius reduction caused by the higher mean molecular weight and the enhanced atmospheric opacity and heat capacity, which slow cooling and contraction. This reversal is absent from the models of \citet{thorngren_bayesian_2018} and \citet{sarkis_evidence_2021}, where increasing $Z$ always reduces the radius, likely because metal enrichment is included only in the planetary interior, whereas the MESA interior models increase the atmospheric opacity with $Z$. This enhanced opacity \citep[see also][]{muller_theoretical_2020} contributes to the lower heating efficiencies inferred here (Fig.~\ref{Fig: Heating efficiency vs flux}).

The direct effect of metallicity on the magnetic field is comparatively small. For $\epsilon_{\rm inner}=0$, both $B_{\rm rms,dyn}$ and $B_{\rm dip,surf}$ increase moderately with $Z$, whereas for heated models $B_{\rm rms,dyn}$ remains nearly independent of metallicity. The increase in planetary radius nevertheless reduces $B_{\rm dip,surf}$ by $\sim20\%$ for the most inflated models ($\epsilon_{\rm inner}=0.2$), with an even weaker effect for $\epsilon_{\rm inner}=0.02$. Thus, the dynamo scaling itself is only weakly sensitive to $Z$. In these experiments, we have assumed $R_{\rm dyn} = r(P=\mathrm{1 \, Mbar})$.

\begin{figure*}[t]
\centering
\includegraphics[width=.9\hsize]{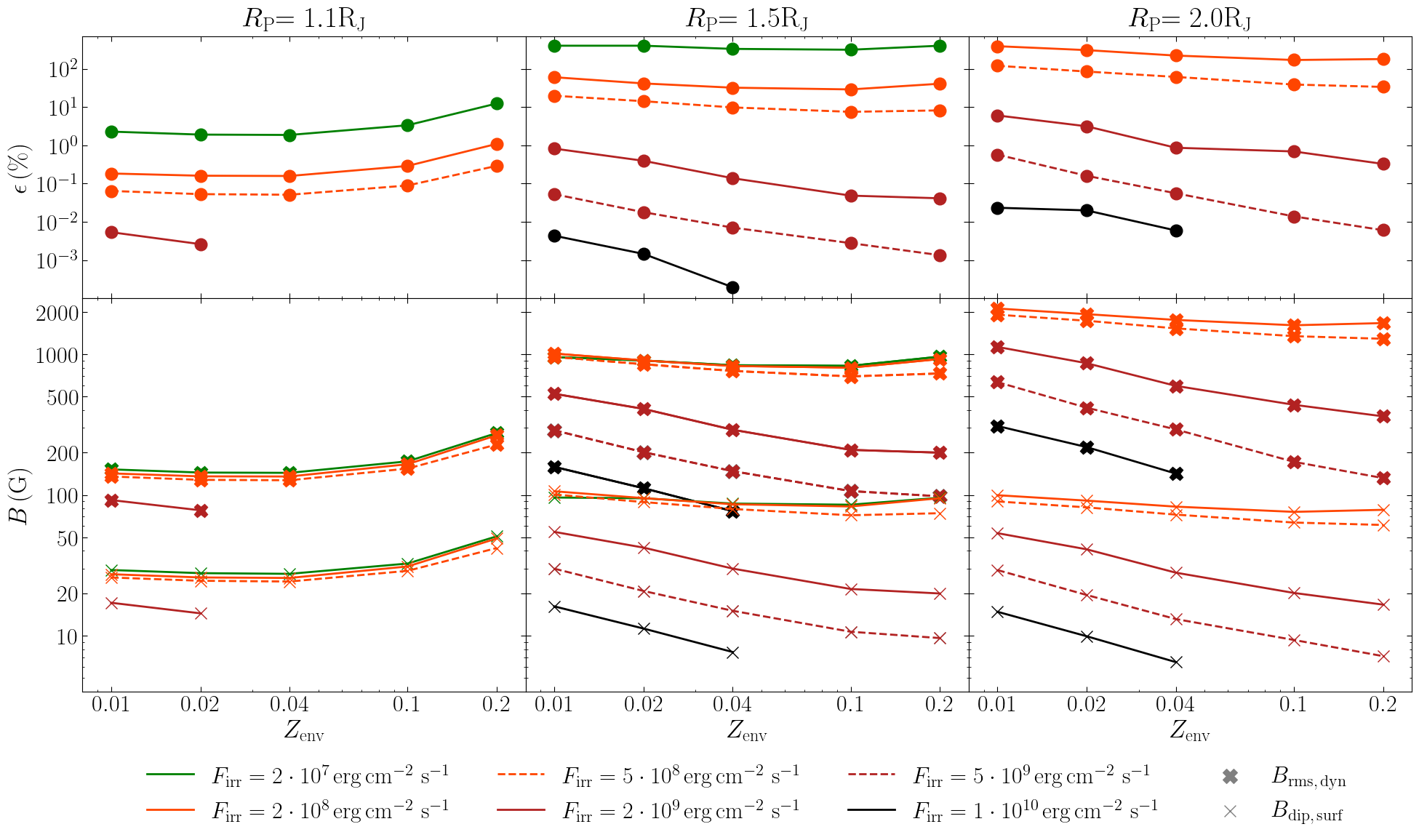}
\caption{Magnetic field estimates obtained using interpolation grids constructed with five different envelope metallicities. Each column corresponds to a hypothetical $1\,\mathrm{M_J}$ planet with an observed radius of 1.1, 1.5, or $2.0\,\mathrm{R_J}$. The upper panels show the heating efficiency required to reproduce the observed radius, while the lower panels show the corresponding magnetic field predictions. Colors denote different irradiation levels. The figure illustrates how uncertainties in the assumed envelope metallicity propagate into the inferred heating efficiencies and, consequently, into the magnetic field estimates.}
\label{Fig: Brms and eps vs Z}
\end{figure*}

Metallicity has a larger effect when interpreting observations because different combinations of $Z$ and heating efficiency can reproduce the same observed radius. This $Z$--$\epsilon$ degeneracy \citep{thorngren_bayesian_2018, sarkis_evidence_2021} therefore propagates into the inferred magnetic fields. To quantify it, we constructed five independent interpolation grids following Sect.~\ref{Sec: Simulation grid and interpolator}, each with a different metallicity. Figure~\ref{Fig: Brms and eps vs Z} shows the inferred $\epsilon$ and magnetic fields for hypothetical $1\,\mathrm{M_J}$ planets with radii of 1.1, 1.5, and $2.0\,\mathrm{R_J}$ over a range of irradiation levels.

For mildly inflated planets ($R_{\rm P}=1.1\,\mathrm{R_J}$), increasing $Z$ requires larger heating efficiencies to compensate for the smaller intrinsic radii of metal-rich planets, resulting in stronger inferred magnetic fields. This trend weakens with inflation and reverses at $R_{\rm P}=2.0\,\mathrm{R_J}$, where the enhanced opacity allows the inflated radius to be maintained with less heating and therefore weaker fields. Across the explored metallicity range, the inferred $\epsilon$ for a $2\,\mathrm{R_J}$ planet spans more than an order of magnitude, while the magnetic field varies by up to a factor of four. Thus, although the dynamo scaling depends only weakly on metallicity, the $Z$--$\epsilon$ degeneracy represents an important systematic uncertainty when applying the interpolation framework to individual HJs. We adopt $Z=0.02$ throughout this work, comparable to but near the lower end of the values inferred by \citet{thorngren_mass-metallicity_2016} and \citet{muller_mass-radius_2024}. If inflated HJs are typically more metal-rich, our inferred magnetic fields may therefore be systematically overestimated.

\section{Error estimation}
\label{App: Error estimation}

\subsection{Observational error propagation}
\label{App: Observational error propagation}

The data from the NASA Exoplanet Archive \citep{christiansen_nasa_2025} generally have asymmetric errors; that is, for any quantity $A$ we have upper and lower errors: $A_{\sigma_{A-}}^{\sigma_{A+}}$. The Monte Carlo propagation (App.~\ref{App: Interpolator error propagation}) samples both; in the analytic expressions below, being conservative, we use the maximum of both limits: $\sigma_{A} = \max(\sigma_{A+},\sigma_{A-})$. 

Some exoplanets lack a specified orbital distance, which we needed to obtain $F_{\rm irr}$. In these cases, we obtain it with Kepler's third law:
\begin{equation*}
    a^3 = \frac{G M}{4 \pi^2} P^2 \quad \rightarrow \quad a({\rm AU}) = \big(k M_{\star}(M_{\odot})P({\rm days})^2\big)^{1/3}
\end{equation*}
where $k = 7.496 \times 10^{-6}$ AU$^3$/days$^2$. Then, its corresponding standard deviation, assuming symmetric errors for orbital period and stellar mass, is:
\begin{equation*}
    \sigma_a = \frac{a}{3} \left[  \left( \frac{\sigma_{M_\star}}{M_\star}\right)^2 + \left( \frac{2\sigma_P}{P}\right)^2 \right]^{1/2}
\end{equation*}

The standard deviation for the received planetary flux, $F_{\rm irr} = \sigma_{\rm SB} T_{\rm eff}^4 \,R_{\star}^2/a^2$, in terms of the other symmetric standard deviations is:
\begin{equation*}
    \sigma_{F_{\rm irr}} = 2 F_{\rm irr} \left[\left(\frac{2 \sigma_T}{T}\right)^2 + \left(\frac{\sigma_{R_\star}}{R_\star}\right)^2 + \left(\frac{\sigma_a}{a}\right)^2 \right]^{1/2}
\end{equation*}
These standard deviations are used to propagate the errors through the interpolators and scaling laws.

\subsection{Interpolator error propagation}
\label{App: Interpolator error propagation}

To propagate the errors through the interpolators, we first considered a simple linear error propagation:
\begin{equation*}
\begin{aligned}
\sigma_A = \left[ \left( \frac{\partial A}{\partial \log_{10}F} \right)^2 \sigma_{\log_{10}F}^2 + \left( \frac{\partial A}{\partial M} \right)^2 \sigma_M^2 \right.\\
\left. + \left( \frac{\partial A}{\partial R} \right)^2 \sigma_R^2 + \left( \frac{\partial A}{\partial P} \right)^2 \sigma_P^2 \right]^{1/2}.
\end{aligned}
\end{equation*}
where $A$ represents $B_{\rm rms, dyn}$, $B_{\rm dip, surf}$, or $\epsilon$ for either heating mechanism (i.e., the six interpolators described in Sect.~\ref{Sec: Simulation grid and interpolator}). Each partial derivative was obtained numerically by calculating slopes with interpolated nearby points. For example, for $M_{\rm P}$:
\begin{equation*} 
\frac{\partial A}{\partial M} = \frac{\mathcal{I}_{A}(F, M + \delta M, R, P) - \mathcal{I}_{A}(F, M - \delta M, R, P)}{2 \delta M}
\end{equation*}
and analogously for $\log_{10}F$, $R$, and $P$. This approach provides a computationally inexpensive estimate of the propagated uncertainty and generally reproduces the standard deviation well. However, assuming symmetric uncertainties becomes problematic for the heating efficiencies, whose large relative errors often produce $1\sigma$ intervals extending to negative values, which are unphysical for $\epsilon$.

To overcome this limitation, we instead adopt a Monte Carlo error propagation. For each planet, we generate $N=10^4$ realizations of the received flux, mass, radius, and orbital period, drawn from split-normal distributions with the lower and upper uncertainties of each quantity (Gaussian for $P_{\rm orb}$). Each realization is then passed through the interpolator; realizations with a radius below the corresponding $R_{\epsilon=0}$ are assigned $\epsilon=0$ and the fields of the non-inflated model. We report the median with uncertainties corresponding to the 16th and 84th percentiles of this distribution, equivalent to $\pm1\sigma$ for a Gaussian distribution while naturally preserving the asymmetry introduced by the interpolation.

To further assess the uncertainty propagation, we also tested a Markov chain Monte Carlo (MCMC) approach using the Python package \texttt{emcee} \citep{foreman-mackey_emcee_2013}, which implements the affine-invariant ensemble sampler of \cite{goodman_ensemble_2010}. We adopted Gaussian priors for $F_{\rm irr}$, $M_{\rm P}$, and $R_{\rm P}$ using the same means and standard deviations described in App.~\ref{App: Observational error propagation}. The resulting uncertainties in the fitted parameters were nearly identical to those obtained with the Monte Carlo sampling described above, but at a substantially higher computational cost. Given that this work aims to provide order-of-magnitude estimates, we adopt the simpler Monte Carlo approach throughout.

In addition to these propagated observational uncertainties, our interpolated results are also subject to systematic uncertainties arising from the underlying model assumptions, including the adopted equation of state, the total heavy-element content and its internal distribution, and the magnetic field scaling laws. These effects are discussed in App.~\ref{App: Interior model caveats}.

\subsection{Scaling law error propagation}
\label{App: Scaling laws error propagation}

After obtaining the interpolated magnetic field estimates, we estimate the uncertainties in the scaling-law coefficients also using a Monte Carlo strategy. As an example, for the MRL fit of the $\epsilon_{\rm inner}$ models we adopt the functional form
\begin{equation*}
    \ln(B_{\rm rms, dyn}(\epsilon_{\rm inner})) = \ln K + \alpha \ln \frac{M_{\rm P}}{\mathrm{M_J}} + \beta \ln \frac{R_{\rm P}}{\mathrm{R_J}} + \gamma \ln \frac{L_{\rm P}}{\mathrm{L_J}}
\end{equation*}
where $L_{\rm P}=\epsilon_{\rm inner}\pi R_{\rm P}^2F_{\rm irr}$. For each planet, we generate random realizations of the planetary mass, radius, and incident stellar flux by sampling split-normal distributions centered on their observed values. An additional systematic uncertainty of a factor of two was assumed for the magnetic field estimates to account for uncertainties in the underlying interior model and the dynamo scaling law (see App.~\ref{App: Interior model caveats}). This uncertainty was added in quadrature to the propagated observational uncertainty in logarithmic space, $(\sigma_{\ln B}^{\rm tot})^2 = (\sigma_{B_{\rm rms,dyn}}/B_{\rm rms,dyn})^2 + (\ln 2)^2$, before generating the Monte Carlo realizations, and therefore the sampled distribution is $\mathcal{N} (\ln B_{\rm rms,dyn} , (\sigma_{\ln B}^{\rm tot})^2)$, i.e., log-normal, spanning a factor of two above and below $B_{\rm rms,dyn}$ at $1\sigma$. For each Monte Carlo realization, we fit the above expression using a nonlinear least-squares algorithm, obtaining one set of coefficients $(K,\alpha,\beta,\gamma)$. Repeating this procedure generates a distribution for each fitted coefficient, from which we adopt the mean values as the best-fit parameters, as well as the upper and lower uncertainties with the 16th and 84th percentiles.

Finally, we also propagate uncertainties when applying the scaling laws to individual planets. Using again the MRL fit for $B_{\rm rms,dyn}(\epsilon_{\rm inner})$ as an example, we generate Monte Carlo realizations of both the planetary parameters and the fitted coefficients. The planetary quantities $M_{\rm P}$ and $R_{\rm P}$ are sampled from split-normal distributions centered on their measured values, and $L_{\rm P}$ (or $F_{\rm irr}$) from a Gaussian distribution, while the fit coefficients ($K$, $\alpha$, $\beta$, and $\gamma$) are sampled from split-Gaussian distributions using the upper and lower uncertainties obtained from the fitting procedure. For each realization, the scaling law is evaluated to obtain one estimate of $B_{\rm rms,dyn}(\epsilon_{\rm inner})$, producing a population of predicted magnetic field strengths. The reported uncertainty is then computed from the 16th and 84th percentiles of this distribution, thereby accounting simultaneously for the uncertainties in the planetary properties and in the scaling-law coefficients.

\end{appendix}

\end{document}